\documentclass[12pt]{article}
\usepackage{multirow}
\usepackage{calc}
\usepackage{amsmath,amssymb,amsthm,amscd}
\numberwithin{equation}{section}
\usepackage[bf]{caption}
\usepackage{longtable}
\usepackage{array}
\usepackage{enumerate}
\usepackage{float}
\usepackage{subfig}
\usepackage{multicol}
\usepackage{multirow}
\usepackage{epsfig}
\usepackage{graphicx}
\usepackage{pict2e}
\usepackage{color}
\usepackage{authblk}
\usepackage{cite}
\usepackage[all]{xy}
\usepackage[width=1.09\textwidth]{caption}
\usepackage{bbm}
\usepackage{youngtab}
\usepackage{hyperref}

\usepackage[utf8]{inputenc}
\usepackage[T1]{fontenc}
\usepackage[english]{babel}
\usepackage{xspace}
\usepackage{pifont}

\numberwithin{equation}{section}

\def\hybrid{\topmargin -20pt    \oddsidemargin 0pt
        \headheight 0pt \headsep 0pt
        \textwidth 6.25in      
        \textheight 9 in      
        \marginparwidth .875in
        \parskip 5pt plus 1pt
          \jot = 1.5ex
  }
\hybrid

\newcommand{\beq}{\begin{equation}}
\newcommand{\eeq}{\end{equation}}

\newcommand{\bea}{\begin{eqnarray}}
\newcommand{\eea}{\end{eqnarray}}

\newcommand{\nn}{\nonumber}

\def\P{\mathbb{P}}

\newcommand{\cS}{\mathcal{S}}

\newcommand{\one}[0]{\ensuremath{\mathbf{1} }\xspace}

\newcommand{\C}{\mathbb{C}}

\newcommand{\bbF}{\mathbb{F}}

\newcommand{\suppress}[1]{}

\newcommand{\cref}{{\bf [check ref]}}

\newcommand{\tr}{\mathrm{Tr}\:}

\def\blfootnote{\xdef\@thefnmark{}\@footnotetext}
\long\def\symbolfootnote[#1]#2{\begingroup%
\def\thefootnote{\fnsymbol{footnote}}\footnote[#1]{#2}\endgroup}

\begin{document}

\baselineskip=15pt

\begin{titlepage}
\begin{flushright}
\parbox[t]{1.8in}{\flushright 
CERN-TH-2016-063\\
MIT-CTP-4773}
\end{flushright}

\begin{center}

\vspace*{ 1.2cm}

{\Large Three-Index Symmetric Matter Representations of SU(2)
 in F-Theory from Non-Tate Form Weierstrass Models}

\vskip 1.2cm

\renewcommand{\thefootnote}{}
\begin{center}
 { Denis Klevers$^1$, Washington
   Taylor$^2$\ \footnote{{\tt denis.klevers}
at {\tt cern.ch}, 
{\tt wati} at {\tt mit.edu}} 
 }
\end{center}
\vspace{0.4cm}
{\it \small
${}^{1}$Theoretical Physics Department, CERN, CH-1211 Geneva 23, Switzerland\\[.3cm]
${}^2$Center for Theoretical Physics, Department of Physics,
Massachusetts Institute of Technology, 77 Massachusetts Avenue
Cambridge, MA 02139, USA
}

 \vspace*{0.3cm}

\end{center}

\vskip 0.2cm
 
\begin{center} {\bf ABSTRACT } \end{center}

We give an explicit construction of a class of F-theory models with
matter in the three-index  symmetric
({\bf 4}) representation of SU(2).  This
matter is realized at codimension two loci in the F-theory base where
the divisor carrying the gauge group is singular; the associated Weierstrass
model does not have the form associated with a generic SU(2) Tate
model.  For 6D theories, the matter is localized at a triple point singularity of
arithmetic genus $g = 3$ in the curve supporting the SU(2) group.
This is the first explicit realization of matter in F-theory in a
representation corresponding to a genus contribution greater than one.
The construction is realized by ``unHiggsing'' a model with a U(1)
gauge factor under which there is matter with charge $q=3$.  The resulting
SU(2) models can be further unHiggsed to realize non-Abelian $G_2
\times \text{SU}(2)$ models with more conventional matter content or
$\text{SU}(2)^3$ models with trifundamental matter.  The U(1) models used as
the basis for this construction do not seem to have a Weierstrass
realization in the general form found by Morrison-Park, suggesting
that a generalization of that form may be needed to incorporate models
with arbitrary matter representations and gauge groups localized on
singular divisors.
~
\\
~\\
\phantom{.}\hfill {April 2016}
\end{titlepage}

\tableofcontents

\section{Introduction}

F-theory \cite{Vafa:1996xn, Morrison:1996na, Morrison:1996pp} provides
a very general string-theoretic approach to constructing low-energy
theories of supergravity coupled to gauge fields and matter.  In
particular, F-theory extends the approach of type IIB string theory to
include non-perturbative seven-brane configurations that produce a rich
variety of structures for low-energy physics.  F-theory uses the
axiodilaton of the IIB theory to encode an elliptic fibration over the
compactification space.  

A beautiful mathematical correspondence
originally elucidated by Kodaira \cite{kodaira1963compact} relates
singularities in the elliptic fibration over (complex) codimension one
subspaces (divisors) in the compactification space to Dynkin diagrams,
encoding the physical non-Abelian gauge content of the theory in
geometric structure.  This correspondence is well-understood, and has
been used to study low-energy theories with exceptional gauge groups
($E_6, E_7, E_8$) and non-simply laced groups (Sp($N$), $F_4, G_2$) in
addition to the usual groups such as SU($N$) that have standard
realizations on D-branes in perturbative string theory.  A similar
correspondence holds between codimension two singularities in elliptic
fibrations and the representation content of matter in F-theory
models, but this correspondence is at present only partially
understood despite much recent work in the F-theory community on the
explicit resolution of codimension two singularities
\cite{Katz:1996xe, Bershadsky:1996nh, Morrison:2011mb, Esole:2011sm,
  Hayashi:2014kca, Esole:2014bka-12, Braun:2014kla, Grassi:2013kha}.
In this paper we explore some explicit examples of F-theory models
with novel matter content as a step towards a more general
understanding of the codimension two generalization of the Kodaira
story.

Some hints towards a general structure underlying the proposed
correspondence between codimension two singularities in elliptic
fibrations and representation theory of semi-simple Lie groups  were
given in \cite{Kumar:2010am, Morrison:2011mb}.  For any representation
$\mathbf{R}$ of a Lie group $G$, 
there is a number $g_R$ given by
\begin{equation}
g_R = \frac{ \lambda}{12}\left(2 \lambda C_{\mathbf{R}} + B_{\mathbf{R}} - A_{\mathbf{R}} \right) \,,
\label{eq:genus-contribution}
\end{equation}
where $A_{\mathbf{R}}, B_{\mathbf{R}}, C_{\mathbf{R}}$ are numerical coefficients associated with the
representation $\mathbf{R}$ through
\begin{align}
\tr_\mathbf{R} F^2 & = A_\mathbf{R}  \tr F^2 \\
\tr_\mathbf{R} F^4 & = B_\mathbf{R} \tr F^4+C_\mathbf{R} (\tr F^2)^2 \,,
\end{align}
and $\lambda$ is a group-dependent constant, with $\lambda = 1$ for SU($N$).
Here $\tr$ refers to the trace in the fundamental representation, while
${\rm Tr}_\mathbf{R}$ corresponds to the trace in the representation $\mathbf{R}$.  By
manipulation of the anomaly cancellation formulae of 6D supergravity,
it was suggested in \cite{Kumar:2010am} that $g_\mathbf{R}$ should have a natural geometric
interpretation as a genus contribution to the divisor (curve)
supporting the gauge group.
Previous analyses of
specific cases have supported this hypothesis.  For SU($N$), $k$-index
antisymmetric representations all have $g_\mathbf{R} = 0$, and these are
precisely the representations that can be realized on a smooth genus 0
curve in a 6D F-theory model.  The adjoint and
(two-index) symmetric matter
representations of SU($N$) both have $g_\mathbf{R} = 1$.  In 6D models where $G$ is
realized on a smooth curve of genus $g$, there are $g$ matter fields
in the adjoint representation.  
We expect that for all representations with $g_\mathbf{R} > 0$ other than the
adjoint, $g_\mathbf{R}$ represents
the \emph{arithmetic genus} contribution from a
singularity $p$ on the divisor $C$ that supports the group $G$, where
$p$ supports matter in the representation $\mathbf{R}$.  

As discussed in general terms in
\cite{Sadov:1996zm, Morrison:2011mb}, the two-index
symmetric representation of SU($N$)  is
expected to be realized on ordinary double point singularities of the
singular curve $C$ carrying the group.  
Recently, two explicit
constructions of classes of models containing matter in the two-index
symmetric representation ({\bf 6})
of SU(3) were given \cite{Cvetic:2015ioa, Anderson:2015cqy}.
Direct construction of Weierstrass models with $g_\mathbf{R} > 0$ matter
representations other than the adjoint appears to be quite subtle, as
the algebraic structure of {\it e.g.} SU($N$) models with such matter
requires an intricate cancellation in the vanishing of the
discriminant to high order on $C$ that relies on the singular nature
of $C$ and the consequent non-UFD (Universal Factorization Domain) structure of the ring of functions
on $C$.
Such models thus cannot be realized 
as Weierstrass forms from generic constructions in the standard Tate approach
used in {\it e.g.} \cite{Bershadsky:1996nh, Katz:2011qp},
or using a naive power series analysis using generic
factorization properties of functions in
$C$ as in \cite{Morrison:2011mb}.
Lacking a general theory of Weierstrass forms for models with such
exotic matter representations,  explicit constructions of symmetric
matter representations have so far
used indirect approaches.  In 
\cite{Cvetic:2015ioa}, the symmetric
representation of SU(3) was constructed by identifying models with
Abelian groups U(1)$\times$U(1) and appropriate charges that lift
to the symmetric representation of SU(3) after unHiggsing.
This is the general approach we use in this paper.
In \cite{Anderson:2015cqy}, the symmetric representation of SU(3) was
identified by Higgsing a theory with a larger (SU(6)) group so that
the symmetric matter naturally appeared after the Higgsing.  This
gives a complementary perspective on the construction of such models
that we also incorporate into the analysis of this paper.
A more direct approach to constructing Weierstrass models for these
kinds of situations where the ring of functions on the  singular divisor $C$
is not a UFD will be presented elsewhere \cite{kmrt}.

In this paper we focus on the three-index symmetric ({\bf 4})
representation of SU(2), associated with the Young diagram
${\tiny\yng(3)}$. We  realize this representation by unHiggsing  Abelian
models constructed in \cite{Klevers:2014bqa} with
U(1) gauge group and matter of charge $q=3$. For SU(2), there is no quartic Casimir, so the
group coefficient $B_\mathbf{R}$ vanishes, and we have $A_\mathbf{4} = 10$, 
$C_\mathbf{4} = 41$ for the $\mathbf{4}$ representation.
These coefficients are readily verified by using a field
strength $F$ proportional to the generator $T_3$, which takes the form
${\rm diag}(1/2, -1/2)$ in the fundamental representation and ${\rm
  diag}(3/2, 1/2, -1/2, -3/2)$ in the three-index
symmetric representation {\bf 4}.  It follows from
(\ref{eq:genus-contribution}) that the genus contribution from a full
hypermultiplet in the {\bf 4} representation of SU(2) is $g_\mathbf{4} =
6$.  Because this representation is self-conjugate (pseudoreal), we
can have matter in a half-hypermultiplet, giving a
genus contribution $\frac{1}{2}g_\mathbf{4} = 3$.  
From the point of view of 6D anomaly cancellation, the contribution of
a half-hypermultiplet  in the {\bf 4} representation combined with
7 hypermultiplets in the fundamental {\bf 2} representation are
\emph{anomaly equivalent} \cite{Morrison:2011mb, Grassi:2011hq} to the
contribution of 3 hypermultiplets in the adjoint {\bf 3}
representation along with 7 uncharged hypermultiplets.
We thus expect that we may find
half-hypermultiplets of the {\bf 4} representation of SU(2) at
arithmetic genus 3 singularities in a curve $C$ supporting the gauge
group in a general complex surface base $B$.  We see that this works out
as expected in the explicit constructions we present here based on unHiggsing the U(1) models
in  \cite{Klevers:2014bqa}.  As in the
previous explicit constructions of symmetric ({\bf 6}) matter
representations of SU(3), the models that we find have a non-Tate
realization of the gauge group SU(2) in the Weierstrass model.  This
matches with the general expectations of the analysis of
\cite{Anderson:2015cqy} and seems to be related to another curious
feature of the construction shown here, which is that the involved U(1) model
of  \cite{Klevers:2014bqa} does not have the general form considered in
\cite{Morrison:2012ei}.  We discuss these connections further in the
conclusions section at the end of the paper.

The structure of this paper is as follows.  In  Section~\ref{sec:AbelianModel} we review 
the U(1) models of  \cite{Klevers:2014bqa} with charge $q = 3$ matter.
In Section~\ref{sec:3-matter}, we unHiggs these U(1) models to
SU(2) models with matter in the {\bf 4} representation.
 In Section~\ref{sec:further-unHiggsing}, we consider further
 unHiggsing to non-Abelian gauge groups with other matter content, and
 Section~\ref{sec:conclusions} contains some concluding remarks.

\section{Abelian F-theory models with matter of charge $q=3$}
\label{sec:AbelianModel}

In this section, we review a construction of a family of F-theory
compactifications with gauge group $G=\text{U}(1)$ and matter with
U(1) charges $q=1,\,2,\,3$.  These compactifications were first
studied in \cite{Klevers:2014bqa}, to which we refer for further
details. In Section \ref{sec:AbelianCY}, we briefly recall the
construction of the elliptically fibered Calabi-Yau manifolds, denoted
by $X$, specifying these compactifications. We then summarize the
matter spectrum of the resulting effective theories in Section
\ref{sec:AbelianMatter}. We conclude this discussion in Section
\ref{sec:AbelianModlesP2} by presenting explicit models with base
$B=\mathbb{P}^2$.

\subsection{Geometry of the elliptic fibration}
\label{sec:AbelianCY}

We consider elliptically fibered Calabi-Yau manifolds
$\pi:X\rightarrow B$ with base manifold $B$.  The elliptic fiber
$\mathcal{E}=\pi^{-1}(p)$ over a generic point $p\in B$ is given by
the Calabi-Yau hypersurface in the del Pezzo surface $dP_1$, which is
the blow-up of $\mathbb{P}^2$ at a point;
this space is also known as the
first Hirzebruch surface $\bbF_1$. F-theory compactifications on such
Calabi-Yau manifolds $X$ were first analyzed in detail in
\cite{Klevers:2014bqa}, whose notation and conventions we follow.  In
summary, the resulting low-energy effective theories have
$G=\text{U}(1)$ gauge group and charged matter with U(1) charges
$q=1,\,2,\,3$.

\begin{figure}[ht]
\centering \def\svgwidth{300pt} 
~\hspace{1.7cm}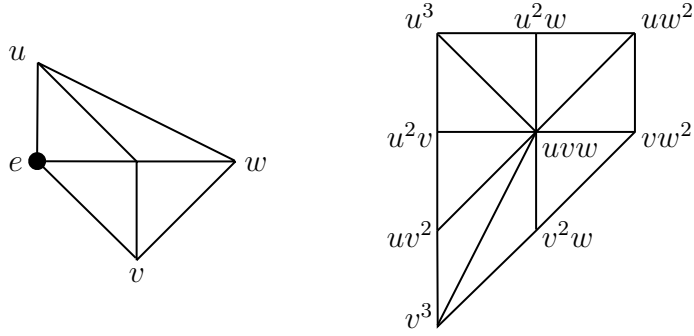
\caption{Polyhedron for $dP_1$ and its dual with corresponding monomials (in the patch $e=1$). The toric zero section $\hat{c}_0$ is indicated by the dot.}
\label{fig:F3}
\end{figure}

The Calabi-Yau manifold $X$ is constructed as the hypersurface
\beq \label{eq:PF3}
	p:=s_1 u^3e^2 + s_2 u^2 ve^2  + s_3 u v^2 e^2  +  s_4 v^3e^2 +
  s_5 u^2 w e
+  s_6 u v w e +  s_7 v^2 w e + s_8 u w^2 + s_9 v w^2=0 \,,
\eeq
in the ambient space of a $dP_1$ fibration over $B$.  Here the
coefficients $s_i$ are sections of line bundles on the base $B$, to be
specified momentarily, and the variables $[u\!:\!v\!:\!w\!:\!e]$ are
the homogeneous coordinates on $dP_1$, which is the ambient space of
the generic elliptic fiber $\mathcal{E}$; the weights of the
coordinates are $(1, 1, 1, 0)$ and $(0, 0, 1, 1)$ with respect to two
$\C^*$ actions on $dP_1$. 
The blow down map from $dP_1$ to $\mathbb{P}^2$ is given by $[u\!:\!v\!:\!w\!:\!e]\mapsto
[ue\!:\!ve\!:\!w]$ so that $e$ vanishes on the exceptional divisor $E$
of $dP_1$.  The del Pezzo surface $dP_1$ is toric; it is described by
a reflexive polyhedron that we depict, along with its dual
polyhedron, in Figure~\ref{fig:F3}.

The Calabi-Yau condition for $X$ implies that  the hypersurface constraint \eqref{eq:PF3} has to be a well-defined section 
of the anti-canonical bundle of the ambient space given by the $dP_1$
fibration over $B$. This requires that the 
coordinates $[u\!:\!v\!:\!w\!:\!e]$ and the coefficients $s_i$ are sections of the following line bundles:
\beq \label{eq:cubicsections}
\text{
\begin{tabular}{c|c}
\text{Section} & \text{Line bundle}\\
\hline
	$u$&$\mathcal{O}(H-E+\cS_9+K_B)$\rule{0pt}{13pt} \\
	$v$&$\mathcal{O}(H-E+\cS_9-\cS_7)$\rule{0pt}{12pt} \\
	$w$&$\mathcal{O}(H)$\rule{0pt}{12pt} \\
	$e_1$& $\mathcal{O}(E)$ \vspace{2cm}\\
\end{tabular}
}\qquad \text{
\begin{tabular}{c|c}
\text{Section} & \text{Line bundle}\\
\hline
	$s_1$&$\mathcal{O}(-3K_B-\cS_7-\cS_9)$\rule{0pt}{13pt} \\
	$s_2$&$\mathcal{O}(-2K_B-\cS_9)$\rule{0pt}{12pt} \\
	$s_3$&$\mathcal{O}(-K_B+\cS_7-\cS_9)$\rule{0pt}{12pt} \\
	$s_4$&$\mathcal{O}(2\cS_7-\cS_9)$\rule{0pt}{12pt} \\
	$s_5$&$\mathcal{O}(-2K_B-\cS_7)$\rule{0pt}{12pt} \\
	$s_6$&$\mathcal{O}(-K_B)$\rule{0pt}{12pt} \\
	$s_7$&$\mathcal{O}(\cS_7)$\rule{0pt}{12pt} \\
	$s_8$&$\mathcal{O}(-K_B+\cS_9-\cS_7)$\rule{0pt}{12pt} \\
	$s_9$&$\mathcal{O}(\cS_9)$ \rule{0pt}{12pt} 
\end{tabular}
} 
\eeq 
Here we denote the line bundle associated to a divisor $D$ by
$\mathcal{O}(D)$, $-K_B$ is the anti-canonical divisor of $B$ and the
classes $H$, $E$ are the pullback of the hyperplane on $\mathbb{P}^2$
and the exceptional divisor on the $dP_1$-fiber, respectively. We note
that the two divisor classes $\mathcal{S}_7$ and $\mathcal{S}_9$,
which are the classes of the coefficients $s_7$ and $s_9$, are free
discrete parameters determining the topology of $X$.
When $\cS_7 =\cS_9 = -K_B$, the $dP_1$ fibration over the base $B$ is
trivial and the  $s_i$ are all sections of the line bundle ${\cal O}
(-K_B)$.  Other values of  $\mathcal{S}_7$ and $\mathcal{S}_9$ parametrize a
two-parameter family of twisted $dP_1$ bundles over $B$.

The Weierstrass model of \eqref{eq:PF3} and a Tate form for it are
readily computed for example using Nagell's algorithm
\cite{Klevers:2014bqa}. As the explicit expressions for the
Weierstrass coefficients $f$, $g$, the discriminant
$\Delta=4f^3+27g^2$ as well as the Tate coefficients are rather
lengthy, we relegate them to \eqref{eq:fgcubic} and
\eqref{eq:Tate} in Appendix \ref{app:RepInWSF}.  The computation of
$\Delta$ reveals that $X$ generically does not exhibit any codimension
one singularities, which implies the absence of a non-Abelian gauge
group in the F-theory effective theory.\footnote{We do not consider the non-Abelian gauge
  groups that would be imposed by choosing bases $B$ with
  non-Higgsable clusters
  \cite{Morrison:2012np,Morrison:2014lca}. However, the analysis
  can be extended straightforwardly.}

The elliptic fibration of $X$ has two sections, one of which being the zero section $\hat{c}_0$ and the second one, 
denoted by $\hat{c}_1$, generating its rank one Mordell-Weil group (MW-group) of rational sections. Consequently, the gauge group $G$ of F-theory on $X$ is
\beq
	G=\text{U}(1)\,.
\eeq
More explicitly, the two sections of $X$ are given by the intersection
of $e=0$ with \eqref{eq:PF3}, which we choose as the zero section
$\hat{c}_0$, and by the second point of intersection of the line
$t_P:=s_8u+s_9v=0$ with $X$, besides $e=0$ where the intersection is
tangent.
 Thus, the MW-group of $X$
is non-toric. In terms of the homogeneous coordinates on the
$dP_1$-fiber, the sections read
\bea \label{eq:s0F3}
   \hat{c}_0\!\!&\!\!=\!\!&\!\! X\cap\{e=0\}:\,\,\, [-s_9:s_8:1:0]\,,\\ 
	\hat{c}_1\!\!&\!\!=\!\!&\!\!X\cap\{t_P=0\} : \,\, [-s_9: s_8:s_1 s_9^3-s_2 s_9^2 s_8+s_3 s_9 s_8^2-s_4 s_8^3:s_7 s_8^2-s_6 s_9 s_8+s_5 s_9^2]\,.\nn
\eea
The Weierstrass coordinates of the section $\hat{c}_1$ are given in
\eqref{eq:WSFQ1F3} in Appendix \ref{app:RepInWSF}, while $\hat{c}_0$
maps to the zero section in Weierstrass form.  The Shioda map of the section $\hat{c}_1$ is
computed to be \cite{Klevers:2014bqa}
\beq \label{eq:ShiodaF3}
	\sigma(\hat{c}_1)=C_1-C_0+3K_B+\cS_7-2\cS_9\,, 
\eeq
where $C_1$, $C_0$ denote the divisor classes of the rational sections
$\hat{c}_1$ and $\hat{c}_0$. The Kaluza-Klein reduction of the M-theory three-form $C_3$ along the $(1,1)$-form associated to $\sigma(\hat{c}_1)$ yields the U(1) gauge field in the effective theory \cite{Morrison:1996pp,Park:2011ji}. The (negative of the) height
pairing is
\beq \label{eq:heightF3}
	 b_{11}=2(-3K_B+2\cS_9-\cS_7)\, ,
\eeq
which enters a Green-Schwarz counterterm in the F-theory effective
action \cite{Park:2011ji, Morrison:2012ei}.

We emphasize here that the locus in $B$ where the coordinates \eqref{eq:s0F3} 
of the two sections agree is given by
\beq \label{eq:def_z1}
z_1:=s_7 s_8^2 - s_6 s_8 s_9 + s_5 s_9^2=0\,.
\eeq
At  points where $z_1 = 0$, a rescaling under the second $\C^*$ makes the two
sections in \eqref{eq:s0F3} 
equivalent.
Note that $z_1$ is precisely the 
$z$-coordinate of $\hat{c}_1$ in  Weierstrass form, \textit{cf.}~\eqref{eq:WSFQ1F3}. 
Thus, the homology class of the divisor in $B$ along which $\hat{c}_0\cong\hat{c}_1$ is  $[z_1]=-2K_B+2\cS_9-\cS_7$
as follows from \eqref{eq:cubicsections}.

Furthermore, we observe 
that the Calabi-Yau constraint \eqref{eq:PF3} is invariant under the $\mathbb{Z}_2$-symmetry $u\leftrightarrow v$ given that we 
also exchange $s_1\leftrightarrow s_4$, $s_2\leftrightarrow s_3$, $s_5\leftrightarrow s_7$ and $s_8\leftrightarrow s_9$. According to \eqref{eq:cubicsections}, this 
amounts to exchanging
\beq \label{eq:Z2symmetry}
	\cS_7\,\mapsto\,\cS_7':=-2K_B-\cS_7 \,,\qquad \cS_9\,\mapsto\, \cS_9':=-K_B+\cS_9-\cS_7\,.
\eeq
This symmetry relates Calabi-Yau manifolds $X$ with the same base $B$, but different values for $\mathcal{S}_7$ and 
$\mathcal{S}_9$. Indeed, we can check that the key geometric properties of $X$ are invariant under the symmetry 
$u\leftrightarrow v$. In particular, this implies that the effective theories of F-theory on $X$ that are related 
by \eqref{eq:Z2symmetry}  have to be identical.

\subsubsection*{Relation to $\text{Bl}_1\mathbb{P}^2(1,1,2)$-elliptic fibrations}

Before delving into the analysis of codimension two singularities of  $X$, we elaborate on
the relation to elliptic fibrations with generic elliptic fiber in $\text{Bl}_1\mathbb{P}^2(1,1,2)$ considered in \cite{Morrison:2012ei}.
We will see that elliptic fibrations with generic elliptic fiber in
$dP_1$ that satisfy the additional condition
$[s_8]=0$ or $[s_9]=0$ are equivalent to those with elliptic fiber in
$\text{Bl}_1\mathbb{P}^2(1,1,2)$.  Indeed, we first note that a
general elliptic fibration $X$ described by \eqref{eq:PF3} has to have
non-vanishing and general coefficients $s_i$. This necessitates that
all divisor classes in \eqref{eq:cubicsections} are effective,
{\it i.e.}~$[s_i]\geq 0$. Second, we see that a model with constant $s_8$
(or $s_9$) allows performing the variable transformation
$u=u'-vs_9/s_8$ ($v=v'-us_8/s_9$) so that we effectively achieve
$s_9\equiv 0$ ($s_8\equiv 0$).\footnote{The symmetry $u\leftrightarrow
  v$ exchanges $s_8\rightarrow s_9$ and the two case of constant $s_8$
  or $s_9$.} As is clear from the dual polyhedron in Figure
\ref{fig:F3}, removing $s_9$ ($s_8$) amounts to blowing up $dP_1$ at
$u=e=0$ ($v=e=0$), {\it i.e.}~adding the vertex with coordinates $(-2,1)$
(or $(-1, -1)$) 
to
the polyhedron of $dP_1$. The resulting polyhedron is precisely the
one of $\text{Bl}_1\mathbb{P}^2(1,1,2)$ and the Calabi-Yau constraint
\eqref{eq:PF3} can be readily written in the form of
\cite{Morrison:2012ei}, as claimed. We will also see this equivalence
on the level of the matter spectrum in Section
\ref{sec:AbelianMatter}.
Note however that, as we discuss in further detail in later sections,
in the generic case where $s_8, s_9 \neq 0$, this class of U(1) models
cannot be written in the Morrison-Park form from
\cite{Morrison:2012ei}.

More extremely, we can relax the effectiveness constraint $[s_8]\geq 0$ or $[s_9]\geq 0$ completely. In both cases, the model 
defined by \eqref{eq:PF3} still defines a sensible elliptically fibered Calabi-Yau manifold. However, there will be a codimension one
singularity of Kodaira type $I_2$ at $s_9=0$ or $s_8=0$, respectively, as analyzed in \cite{Anderson:2014yva,Klevers:2014bqa}. It 
can be resolved globally by the blow-ups in $dP_1$ at $v=e=0$ or $u=e=0$, respectively, resulting again in the new ambient space $\text{Bl}_1\mathbb{P}^2(1,1,2)$. Thus, we see that the elliptic fibrations with their generic elliptic fibers in $\text{Bl}_1\mathbb{P}^2(1,1,2)$ can be thought of as arising from the Calabi-Yau manifold $X$ via the specialization $s_8=0$ 
or $s_9=0$, respectively, in \eqref{eq:PF3}.  

\subsection{The matter spectrum}
\label{sec:AbelianMatter}

The matter spectrum of the F-theory compactification on $X$ is derived
by analyzing the  singularities of the elliptic fibration
that arise over codimension two loci in the base.
Since the Calabi-Yau manifold $X$ has a non-trivial MW-group generated by $\hat{c}_1$,
it automatically has Kodaira fibers of type $I_2$ at the codimension two loci in $B$ along which 
\beq \label{eq:F3charge1}
	y_{1}=f z_1^4+3 x_1^2=0\,
\eeq
is satisfied  \cite{Morrison:2012ei}. 
Here $f$ and $g$ enter the Weierstrass form
of $X$ and $[y_1\!:\!x_1\!:\!z_1]$ are the Weierstrass coordinates of $\hat{c}_1$ given in \eqref{eq:fgcubic} and \eqref{eq:WSFQ1F3}, respectively.
The matter located at \eqref{eq:F3charge1} is automatically charged under 
the  U(1) gauge field corresponding to $\hat{c}_1$. 

The locus \eqref{eq:F3charge1} 
is reducible with three irreducible components, as \textit{e.g.}~shown by a primary decomposition (see
\cite{Cvetic:2013nia,Cvetic:2013uta} for more details on the necessary technical tools), corresponding
to three different matter representations.
The full matter spectrum  derived
in \cite{Klevers:2014bqa} is given in Table~\ref{tab:poly3_matter}, which includes the U(1)-charges, the 
multiplicities $x_{\mathbf{R}}$ of 6D
charged hyper multiplets in the representation $\mathbf{R}$
and the codimension two loci supporting the respective fibers. Here, 
we use the notation $V(I)$ for the vanishing set of an ideal $I$. 

\begin{table}[ht!]
\begin{center}
\footnotesize
\begin{tabular}{|c|c|c|}
\hline
Rep & Multiplicity  & Locus  \\ \hline
$\one_{3}$ & $x_{\mathbf{1}_3}=\cS_9\cdot (-K_B+\cS_9-\cS_7)$  & $V(I_{(3)}):=\{s_8 = s_9 = 0\}$  \rule{0cm}{0.4cm} \\[0.1cm] \hline

$\one_{2}$ & $\begin{array}{c}
 x_{\mathbf{1}_2}=6 K_B^2 \!-K_B\cdot (4\cS_9\!-\!5  \cS_7 )\\
 + \cS_7^2  + 2 \cS_7 \cS_9 - 2 \cS_9^2
\end{array} \!\! $
 &\  $\begin{array}{c}
 V(I_{(2)}):= \{s_4 s_8^3\!-\! s_3 s_8^2 s_9 \!+\! s_2 s_8 s_9^2\! -\! s_1 s_9^3\\
 =s_7 s_8^2  + s_5 s_9^2\!-\! s_6 s_8 s_9 =0\}\backslash\ V(I_{(3)})
\end{array} \rule{0cm}{0.65cm} $
 \\[0.3cm] \hline
$\one_{1}$  &
$\begin{array}{c}
 x_{\mathbf{1}_1}=12 K_B^2 -K_B\cdot (8 \cS_7\!-\! \cS_9)\\
 - 4 \cS_7^2  + \cS_7 \cS_9 - \cS_9^2
\end{array}  $ & $\begin{array}{c}
V(I_{(1)}):=\{\eqref{eq:F3charge1}\}\backslash\ (V(I_{(2)})\cup V(I_{(3)}))\\
~
\end{array}$\rule{0cm}{0.65cm}\\[0.3cm] \hline
\end{tabular}
\caption{\label{tab:poly3_matter} Charged matter under U(1) and
codimension two fibers of $X$.}
\end{center}
\end{table}

The matter spectrum of $X$ is completed by the number of  neutral
hyper multiplets $H_{\text{neut}}$. It has been computed in 
\cite{Klevers:2014bqa} to be
\bea \label{eq:HneutF3}
	H_{\text{neutral}}&=& 13 + 11 K_B^2 + K_B\cdot (3\cS_7+ 4\cS_9) + 3 \cS_7^2  - 2 \cS_7\cdot \cS_9 + 2 \cS_9^2\,.
\eea
Employing this, together with the  charged spectrum in
Table~\ref{tab:poly3_matter},  anomaly-freedom of the 6D U(1) SUGRA
theory is readily checked, following the general prescription  
of \cite{Erler:1993zy,Park:2011wv}.
We note that the matter spectra in Table \ref{tab:poly3_matter} and in \eqref{eq:HneutF3} are invariant under 
the $\mathbb{Z}_2$-symmetry \eqref{eq:Z2symmetry} of $X$.

We stress that one main difference of the matter spectrum  in Table \ref{tab:poly3_matter} and the one of 
$\text{Bl}_1\mathbb{P}^{2}(1,1,2)$-elliptic fibrations studied in \cite{Morrison:2012ei} is the presence of matter fields with $q=3$.  
In turn, it is expected that models without these matter fields should be already described by the models in \cite{Morrison:2012ei}.
Indeed, employing the discussion at the end of the previous section, Calabi-Yau  manifolds $X$ with $x_{\mathbf{1}_3}=0$, which 
requires either $[s_8]=0$ or $[s_9]=0$, are geometrically completely equivalent to 
$\text{Bl}_1\mathbb{P}^{2}(1,1,2)$-elliptic fibrations and so are the effective theories, 
as expected.

\subsection{Models over $B=\mathbb{P}^2$}
\label{sec:AbelianModlesP2}

\begin{figure}[htb] 
\begin{center} \def\svgwidth{300pt} 
~\hspace{2.0cm}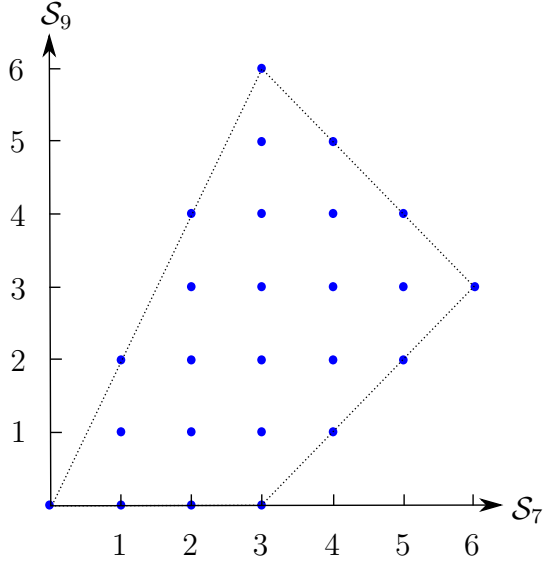 
\caption{Allowed region for the pair $(\cS_7,\cS_9)$ specifying $X$ for $B=\mathbb{P}^2$.} 
\label{fig:allowedregion}
\end{center}
\end{figure}

We conclude the discussion of F-theory compactified on the Calabi-Yau
manifold $X$ by considering the concrete examples with base
$B=\mathbb{P}^2$. In this case we have
$-K_{B}=\mathcal{O}_{\mathbb{P}^2}(3)$ and $\cS_7$ and $\cS_9$ 
can be associated with
non-negative 
integers since the second homology of $\mathbb{P}^2$ is
one-dimensional and generated by the hyperplane $H_B$ of
$\mathbb{P}^2$. We can then solve the conditions imposed by
effectiveness of the divisor classes $[s_i]\geq 0$, $i=1,\ldots, 9$,
given in \eqref{eq:cubicsections}, as in \cite{Cvetic:2013nia}. This
yields the allowed region for the pair $(\cS_7,\cS_9)$ shown in Figure
\ref{fig:allowedregion}.
We immediately notice that this region is precisely given by the toric polytope of $dP_1$ rescaled by $3$, which is precisely the 
anti-canonical class of $\mathbb{P}^2$ in units of $H_B$.

Next we determine the matter spectrum of $X$ for the concrete base $\mathbb{P}^2$ employing Table \ref{tab:poly3_matter}. 
We recall the $\mathbb{Z}_2$-symmetry \eqref{eq:Z2symmetry} relating Calabi-Yau manifolds $X$ with different values for 
$(\cS_7,\cS_9)$. In the allowed region in Figure \ref{fig:allowedregion}, this symmetry exchanges points on the lines $\cS_9=x$ 
and $\cS_9=\cS_7-3+x$ for $x=0,\ldots, 6$. As the effective theories of F-theory on $X$ are related accordingly, as 
discussed before,  and as $\cS_7=3$ is the fixed line under \eqref{eq:Z2symmetry}, we only have to list  models and corresponding spectra for $\cS_7\leq 3$. We obtain the following list for the degrees of the sections $s_i$ entering the Calabi-Yau constraint \eqref{eq:PF3} and of the matter multiplicities $x_\mathbf{R}$:
\beq\label{eq:spectrumP2}
\text{
\begin{tabular}{c||c|c|c|c|c|c|c||c}
$(\cS_7,\cS_9)$ &$[s_1]$&$[s_2]$&$[s_3]$&$[s_4]$&$[s_5]$&$[s_6]$&$[s_8]$& $(x_{\mathbf{1}_3},x_{\mathbf{1}_2},x_{\mathbf{1}_1})$\\
\hline
	$(0,0)$&9&6&3&0&6&3&3&$(0,54,108)$\rule{0pt}{13pt} \\
	$(1,0)$&8&6&4&2&5&3&2&$(0,40,128)$\rule{0pt}{12pt} \\
	$(2,0)$&7&6&5&4&4&3&1&$(0,28,140)$\rule{0pt}{12pt} \\
	$(3,0)$&6&6&6&6&3&3&0&$(0,18,144)$\rule{0pt}{12pt} \\
	$(1,1)$&7&5&3&1&5&3&3&$(3,52,125)$\rule{0pt}{12pt} \\
	$(2,1)$&6&5&4&3&4&3&2&$(2,42,138)$\rule{0pt}{12pt} \\
	$(3,1)$&5&5&5&5&3&3&1&$(1,34,143)$\rule{0pt}{12pt} \\
	$(1,2)$&6&4&2&0&5&3&4&$(8,60,120)$\rule{0pt}{12pt} \\
	$(2,2)$&5&4&3&2&4&3&3&$(6,52,134)$ \rule{0pt}{12pt} \\
    $(3,2)$&4&4&4&4&3&3&2&$(4,46,140)$ \rule{0pt}{12pt} \\
	$(2,3)$&4&3&2&1&4&3&4&$(12,58,128)$ \rule{0pt}{12pt} \\
	$(3,3)$&3&3&3&3&3&3&3&$(9,54,135)$ \rule{0pt}{12pt} \\
	$(2,4)$&3&2&1&0&4&3&5&$(20,60,120)$ \rule{0pt}{12pt} \\
	$(3,4)$&2&2&2&2&3&3&4&$(16,58,128)$ \rule{0pt}{12pt} \\
	$(3,5)$&1&1&1&4&3&3&5&$(25,58,119)$ \rule{0pt}{12pt} \\
	$(3,6)$&0&0&0&0&3&3&6&$(36,54,108)$ \rule{0pt}{12pt} \\
\end{tabular}
}
\eeq
The spectrum of the remaining theories in the allowed region in Figure \ref{fig:allowedregion} can be obtained by application of the $\mathbb{Z}_2$-symmetry \eqref{eq:Z2symmetry}. We note that all the spectra in \eqref{eq:spectrumP2} are different. In particular,
the number of matter fields  with charge $q=2$ is always larger than zero, which will be important for the unHiggsing of $X$ discussed next.

We conclude by noting that the four models with $x_{\mathbf{1}_3}=0$
are precisely four of the possible seven
$\text{Bl}_1\mathbb{P}^2(1,1,2)$-elliptic fibrations that can be
constructed on $B=\mathbb{P}^2$ and without an $I_2$ singularity at
codimension one, {\it i.e.}~an extra SU(2) gauge group factor.  The
role of the parameter $b$ in \cite{Morrison:2012ei,Morrison:2014era}
is played by $b\equiv s_5$, which assumes values from $[b]=3,\ldots,
6$ in the allowed region. In order to obtain the remaining three
models with $[b]=[s_5]=0,1,2$, we have to relax effectiveness of the
class $[s_8]$.  The three missing models are then given at
$(\cS_7,\cS_9)=(4,0),\,(5,0),\,(6,0)$.

\section{Matter in the
three-index symmetric representation $\mathbf{4}$ of SU(2)}
\label{sec:3-matter}

We begin this section by briefly recalling the general geometrical
procedure that corresponds to an unHiggsing of a U(1) to a
non-Abelian gauge symmetry in F-theory. We will focus on unHiggsings
that preserve the rank of the gauge group. General discussions of
rank-preserving unHiggsings of U(1)'s in F-theory can be found in
\cite{Morrison:2012ei,Morrison:2014era,Klevers:2014bqa,Cvetic:2015ioa,Grimm:2015wda}.

An F-theory compactification with a U$(1)^m$ gauge symmetry is
specified by a Calabi-Yau manifold $X_{n+1}^{(m)}$ with a MW-group of
rank $m$. The Abelian gauge symmetry of the theory is unHiggsed to a
non-Abelian one by performing a geometric transition from
$X_{n+1}^{(m)}$ to a new Calabi-Yau manifold $X_{n+1}^{(0)}$ with a
trivial MW-group; the manifold $X_{n+1}^{(0)}$ is obtained by tuning
the complex structure of $X_{n+1}^{(m)}$ so that all its rational
sections are placed on top of each other. Typically, this process
induces codimension one singularities of the elliptic fibration of
$X_{n+1}^{(0)}$ that produce a non-Abelian gauge group in the final
``unHiggsed'' theory. 
This can be thought of as a transition that takes ``horizontal''
divisors in the Calabi-Yau manifold associated with sections into
``vertical'' divisors associated with resolved Kodaira singularities
over divisors in the base.
 For example, it is shown in
\cite{Morrison:2012ei, Morrison:2014era} that a model with a single U(1) gauge group
can be unHiggsed to a model with SU(2) or larger non-Abelian gauge
group\footnote{In some cases, particularly when there are additional
  non-Abelian factors present before the unHiggsing, the unHiggsed
  model can develop problematic singularities.} and the general
unHiggsings of two 
or more U(1)'s are studied in \cite{Cvetic:2015ioa}. Concrete
unHiggsings of toric models with up to three U(1)'s and of general
U$(1)\times$U(1) F-theory compactification are discussed in
\cite{Klevers:2014bqa} and \cite{Cvetic:2015ioa}.

In this section, we analyze the unHiggsing of the Abelian F-theory
model defined by the Calabi-Yau manifold $X$ in \eqref{eq:PF3} that
has one U(1). We thus identify $X^{(1)}\equiv X$.  This model
unHiggses to a non-Abelian theory with $G=\text{SU}(2)$ gauge group,
similar to the models in \cite{Morrison:2012ei,Morrison:2014era}. The
corresponding geometrical tuning of $X$ to a manifold $X^{(0)}\equiv
X^{\text{SU}(2)}$ with trivial MW-group but $I_2$ singularities at
codimension one is discussed in Section \ref{sec:GeoUnHiggsing}.
Then, we show that the structure of codimension two singularities
in $X$ that is responsible for the presence of matter fields with
U(1)-charge $q=3$ in F-theory yields a novel singularity structure in
the unHiggsed geometry $X^{\text{SU}(2)}$: the
$I_2$ singularities corresponding to the SU(2) gauge group occur on a
singular divisor $t=0$ with a triple point singularity. Most notably,
it seems that the triple point singularity can not be deformed without
affecting the $I_2$ singularity of the elliptic fibration of
$X^{\text{SU}(2)}$. This interplay between singularities of the
divisor $t=0$ and the singularity of the elliptic fibration yields a
new non-Tate Weierstrass model with $I_2$ singularities at codimension
one.  Furthermore, as demonstrated in Section \ref{sec:NonAbelMatter},
F-theory on $X^{\text{SU}(2)}$ yields the first explicit realization
of SU(2) gauge theories with the three-index symmetric representation,
which is located precisely at the triple point singularity of the
SU(2) divisor $t=0$. We support this observation by matching the
effective theories before and after the Higgsing in Section
\ref{sec:MatchingHiggsing}. We conclude our discussion by explicitly
constructing all elliptic fibrations $X^{\text{SU}(2)}$ with base
$B=\mathbb{P}^2$.

\subsection{UnHiggsing U$(1)\rightarrow \text{SU}(2)$ in geometry}
\label{sec:GeoUnHiggsing}

\begin{figure}[t]
\def\svgwidth{206pt} 
\centering
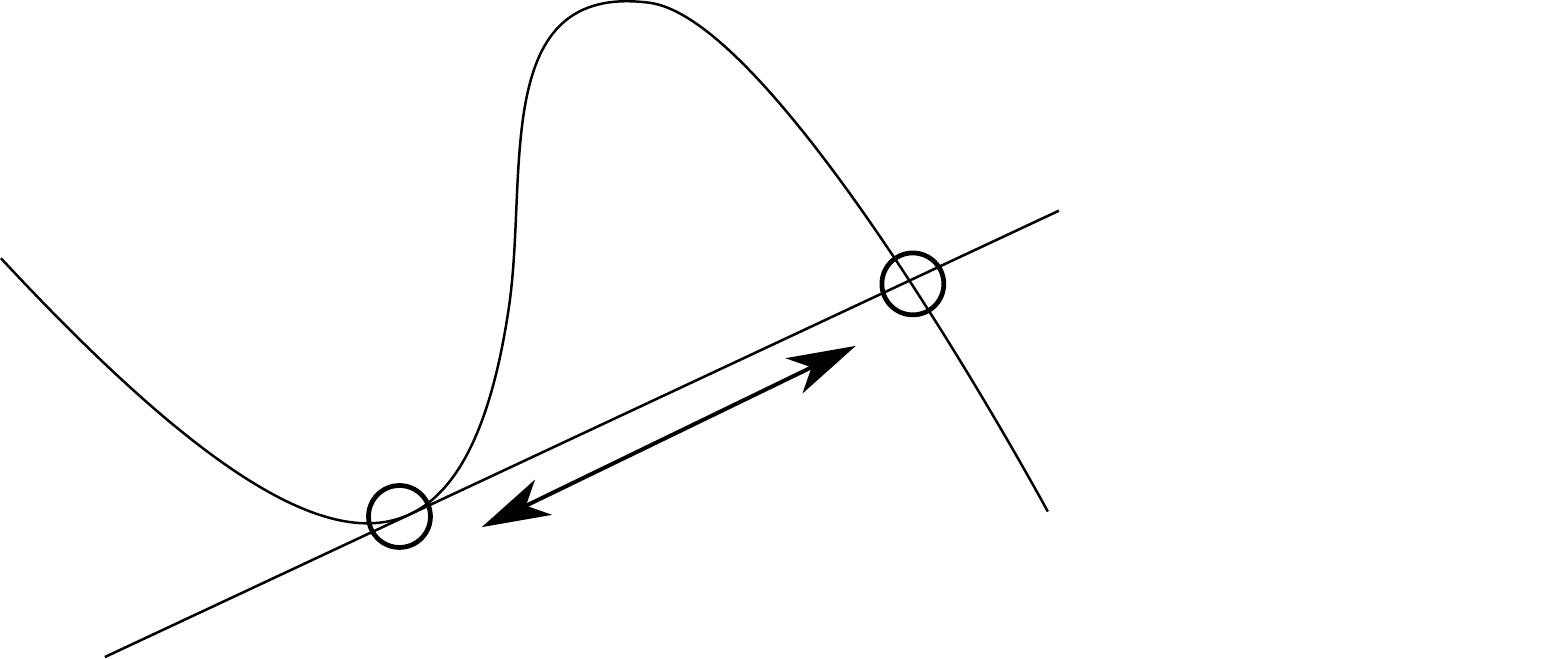\hspace{1cm} \def\svgwidth{174pt} 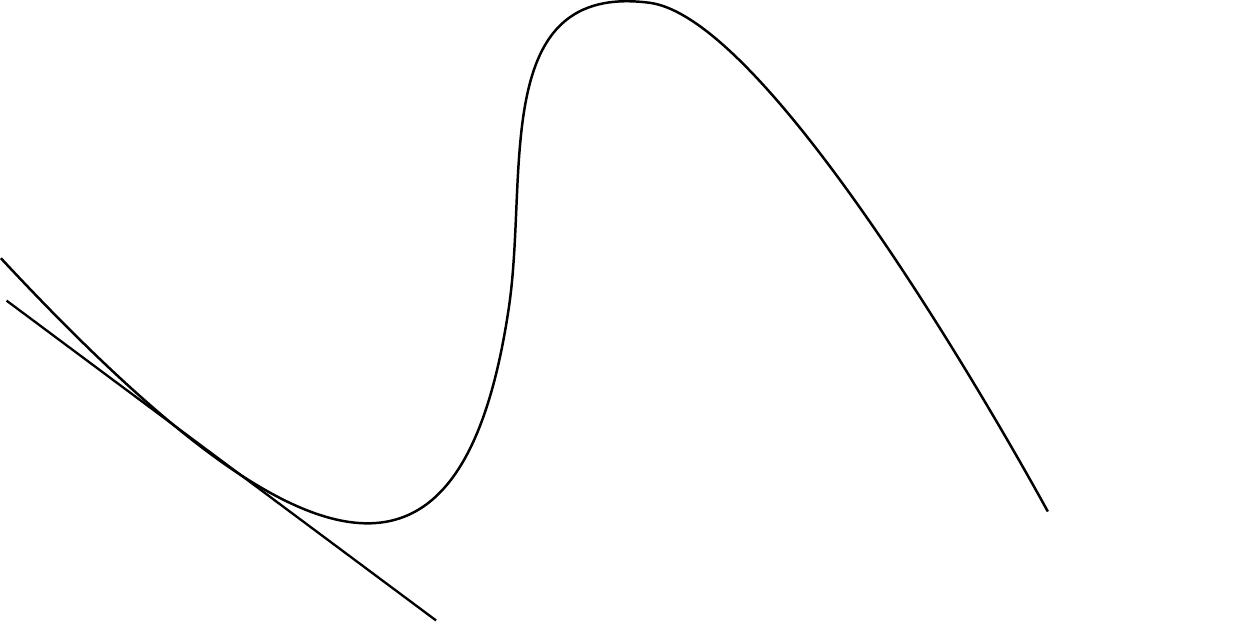
\caption{UnHiggsing by tuning the complex structure of $X$, shown on the left, so that $\hat{c}_0=\hat{c}_1$ in the generic elliptic fiber $\mathcal{E}$ of $X$ as shown on the right.}
\label{fig:unHiggsing}
\end{figure}

We begin by recalling that  the elliptically fibered Calabi-Yau manifold $X$ given in 
\eqref{eq:PF3} has two rational sections $\hat{c}_0$  and $\hat{c}_1$ with fiber coordinates \eqref{eq:s0F3}. The unHiggsing of the 
U(1) gauge symmetry of F-theory on $X$  is performed 
by tuning its complex structure so that the two rational sections $\hat{c}_0$ 
and $\hat{c}_1$ of the elliptic fibration become identical, {\it i.e.}~$\hat{c}_0\equiv \hat{c}_1$, as shown in Figure \ref{fig:unHiggsing}.
As discussed before in \eqref{eq:def_z1}, these two sections coincide precisely if $z_1\equiv 0$,  where $z_1$ is
the $z$-coordinate of the section $\hat{c}_1$ in Weierstrass form.  Thus,  the relevant tuning of the complex structure of $X$ is given by
\beq \label{eq:unHiggs}
	z_1=s_7s_8^2-s_6s_8s_9+s_5s_9^2\rightarrow 0\,.
\eeq
We denote the resulting tuned Calabi-Yau manifold by $X^{\text{SU}(2)}$ for reasons that become clear below.

There are a number of remarks in order.   First, we emphasize that we have to forbid the special solution $s_8=s_9\equiv 0$ to 
\eqref{eq:unHiggs}. This is clear from Table \ref{tab:poly3_matter} because there is matter with $q=3$ located at this locus in $X$. 
This implies that  imposing $s_8=s_9\equiv 0$ globally by  tuning the complex structure of $X$ would render the resulting elliptic fibration of $X^{\text{SU}(2)}$ singular
everywhere, which does not define a good F-theory model. In fact, we consider solutions to \eqref{eq:unHiggs} with
general $s_8$, $s_9$ in order to preserve in the unHiggsing to $X^{\text{SU}(2)}$ the geometric structure in $X$ giving rise to matter with charge $q=3$.

Second, the tuning \eqref{eq:unHiggs} induces a codimension one
singularity of Kodaira type $I_2$. This is immediately clear from
Table \ref{tab:poly3_matter} and can be checked formally for example
by using the Weierstrass form, see Appendix \ref{app:RepInWSF}.
Indeed, the locus $V(I_{(2)})$ in Table \ref{tab:poly3_matter}, which
supports codimension two $I_2$ singularities corresponding to matter
with charge $q=2$, is promoted to codimension one in $B$ if we perform
the tuning $z_1\rightarrow 0$. The locus of $I_2$ singularities is
then given by \beq \label{eq:SU2divisor} t:=s_4 s_8^3\!-\! s_3 s_8^2
s_9 \!+\! s_2 s_8 s_9^2\! -\! s_1 s_9^3=0\,, \eeq whose class is
$[t]=[s_1]+3[s_9]=-3K_B+2\cS_9-\cS_7$ according to
\eqref{eq:cubicsections}.  

Thus, we see that the gauge group $G$ of F-theory on $X^{\text{SU}(2)}$, which has a trivial MW-group of rational
sections, is given by
\beq \label{eq:GGunHiggs}
	G=\text{SU}(2)\,.
\eeq
The U(1) gauge group of $X$ has been unHiggsed in a rank-preserving way to SU(2).

Third, we point out that generically, if all $s_i$ in $z_1$ are
non-trivial and general polynomials, the tuning \eqref{eq:unHiggs}
sets a non-trivial polynomial on $B$ to zero. 
A general solution to this relation can be identified when the base is
smooth (which we assume) and the corresponding ring of sections can be treated as a UFD,
for example when the base is $B = \P^2$, where the sections are simply
homogeneous polynomials of various degrees in the homogeneous
coordinates.
In this case, for example, every factor in $s_9$ must be a factor of
either $s_7$ or $s_8$.  We assume that $s_8$ and $s_9$ have no common
factors since they could be factored out of $z_1$, and
as mentioned above the solution $s_8 = s_9 \equiv 0$
does not give a good F-theory model.
The general solution to \eqref{eq:unHiggs}
with relatively prime $s_8$ and $s_9$ is then given by (cf. \cite{Lawrie:2014uya})
\beq
 \label{eq:UFDSolution} s_5=s_8\sigma_5\,,\qquad
s_6=s_8\sigma_7+s_9\sigma_5\,,\qquad s_7=s_9 \sigma_7\,.  
\eeq
Here
$\sigma_5$ and $\sigma_7$ are arbitrary sections of
$\mathcal{O}(-K_B-\cS_9)$ and $\mathcal{O}(\cS_7-\cS_9)$, as follows
from \eqref{eq:cubicsections}.  Clearly, a necessary condition for the
existence of this solution is effectiveness of
$[s_5]-[s_8]=-K_B-\mathcal{S}_9$ and $\mathcal{S}_7-\mathcal{S}_9$ for
the sections $\sigma_5$ and $\sigma_7$ to exist, respectively.

The constraint \eqref{eq:unHiggs} can also be solved
simply by
setting
\beq 
\label{eq:simpleSol} s_5=s_6=s_7\equiv 0\,. 
\eeq 
Note that this is a special case of (\ref{eq:UFDSolution}), where
$\sigma_5 = \sigma_7 = 0$, and  does not require effectiveness of 
$-K_B-\mathcal{S}_9$  or $\mathcal{S}_7-\mathcal{S}_9$.
We
emphasize that this tuning is clearly always possible on any base $B$. 
The charged matter spectrum of F-theory on
$X^{\text{SU}(2)}$ obtained by this tuning agrees with that obtained
by the tuning \eqref{eq:UFDSolution}. This follows from consistency with the Higgsing back to
$X$ together with the fact, which we checked in an explicit computation, that the additional tuning $\sigma_5=\sigma_7\equiv 0$
does not change the singularities of $X^{\text{SU}(2)}$.
Thus, we will
for the remainder of this work consider the solution
\eqref{eq:simpleSol}.  Finally, we note that simple tunings achieving
$z_1\rightarrow 0$ are possible if $s_8$ or $s_9$ are constants, {\it
  i.e.}~in the absence of matter with U(1)-charge $q=3$, {\it cf.}~Table
\ref{tab:poly3_matter}; for example, if $s_8$ is constant, we can
always solve \eqref{eq:unHiggs} by
$s_7=\frac{1}{s_8^2}(s_6s_8s_9-s_5s_9^2)$.

Let us further elaborate on the geometry of $X^{\text{SU}(2)}$.
First, we emphasize that the divisor $t=0$ defined in
\eqref{eq:SU2divisor} has triple point singularities at the locus of
points defined by $s_8=s_9=0$; {\it i.e.}, three of its branches cross
at the common locus $s_8=s_9=0$. Focusing on complex two-dimensional
bases $B$, $t=0$ defines a Riemann surface with arithmetic genus $g$
computed as
\beq \label{eq:arithmeticGgeneral}
	g=1+\tfrac12 [t]\cdot ([t]+K_B)=p_g+\tfrac12\sum_p m_p(m_p-1)\,.
\eeq
Here the first equality follows from adjunction whereas in the second
equality we split the arithmetic genus into the geometric genus $p_g$
and contributions from all singular points $p$ of $t=0$ with
multiplicity $m_p$, see \textit{e.g.}~\cite{Morrison:2011mb}.  
Each triple point singularity of $t=0$ has
multiplicity $m_p=3$ and contributes $3$ to the arithmetic genus $g$
of $t$ as it can be deformed into three ordinary double point
singularities, each of which contributes one to $g$.  We will discuss
the physical interpretation of the triple point singularity in Section
\ref{sec:NonAbelMatter}, where we show that each triple point
singularity supports a half-hypermultiplet of matter in the
three-index symmetric {\bf 4} representation of SU(2).

We conclude by noting that the geometric genus $p_g$ of the curve
$t=0$ is greater or equal to one for effective classes of $s_8$ and
$s_9$. This follows from the genus formula
\eqref{eq:arithmeticGgeneral} as
\bea \label{eq:geometricGenus}
	p_g\!\!&\!\!=\!\!&\!\!1+\tfrac{1}{2}(-2K_B+[s_8]+[s_9])\cdot(-K_B+[s_8]+[s_9])-3[s_8][s_9]\nn\\\!\!&\!\!\geq\!\!&\!\! 1+\tfrac12 3[s_9]\cdot(-K_B+[s_8]+[s_9])-3[s_8]\cdot[s_9]=1+\tfrac12 3[s_9]\cdot(-K_B+[s_9])-\tfrac32[s_8]\cdot[s_9]\nn\\
	\!\!&\!\!\geq\!\! &\!\!1+\tfrac12 3[s_9]\cdot [s_8]-\tfrac32[s_8]\cdot[s_9] =1\,,
\eea
where we used, employing \eqref{eq:cubicsections}, that
$[t]=-2K_B+[s_8]+[s_9]$ in the first equality, then $-2K_B\geq
2[s_9]-[s_8]$ following from $[s_1]\geq 0$ in the first inequality
and $-K_B+[s_9]\geq [s_8]$ as follows from $[s_7]\geq 0$ in the last
inequality.  Field theoretically, this 
is relevant since we expect the geometric genus to give rise to $p_g$
nonlocal adjoint matter fields.  At least one adjoint matter field
is required to Higgs the SU(2)
gauge theory specified by $X^{\text{SU}(2)}$
back to the original U(1) theory, so if the triple point
singularities do not support localized adjoint matter then it is clear
that the geometric genus of $t$
must be positive.  In addition, we
emphasize that $g\geq 1$ is equivalent to $[z_1]\geq 0$ as we have the
relation \beq [t]=[z_1]-K_B\,, \eeq which follows from
\eqref{eq:def_z1} and \eqref{eq:cubicsections}. This implies that
$[t]$ is always effective as we have $-K_B \geq 0$ and $[z_1]\geq 0$,
which is necessary for the existence of a non-trivial section $z_1$
allowing for the deformation of the model $X^{\text{SU}(2)}$ back to
$X$.  The $p_g$ adjoint Higgs VEV's 
can  be thought of as
corresponding to the deformations in
$z_1\neq 0$.

\subsection{Novel matter structure from non-Tate Weierstrass forms}
\label{sec:WSFI2}

The Weierstrass model of the unHiggsed theory $X^{\text{SU}(2)}$ is 
obtained using the tuning \eqref{eq:simpleSol} in the general
Weierstrass model of $X$ given in \eqref{eq:fgcubic}. 
The resulting SU(2) model is specified by the Weierstrass coefficients  
\bea \label{eq:WSFSU2}
	f&=&\frac{1}{3} \left(-\left(s_3^2-3 s_2 s_4\right) s_8^2+\left(s_2 s_3-9 s_1 s_4\right) s_9 s_8-\left(s_2^2-3 s_1 s_3\right) s_9^2\right)\,,\\
	g&=&\frac{1}{27} \left(-2 (s_3^3 - 9 s_2 s_3 s_4 + 27 s_1 s_4^2) s_8^3 - 
 6 (s_2 s_3^2 + 3 s_2^2 s_4 - 9 s_1 s_3 s_4) s_8^2 s_9\right.
\nonumber\\
 &&\left. + 
 6 s_3 (2 s_2^2 - 3 s_1 s_3) s_8 s_9^2 - 2 s_2^3 s_9^3\right)
 +(s_1 s_4-\tfrac13 s_2 s_3 )T\,.\nn
\eea
Here we have replaced the variable $t$ defined in
\eqref{eq:SU2divisor} for the moment by the formal variable $T$.
While the formal expansion of $f$ and $g$ is thus ambiguous,  it is
clear that $f$ is not naturally written in a form containing terms
proportional to $T$ as there are no cubic terms in
$s_8, s_9$, and this form of $g$ is a fairly natural way of combining
terms with a term linear in $T$.  Alternative presentations of $g$
lead to equivalent conclusions but with different algebra.
From \eqref{eq:WSFSU2}, we readily compute the discriminant
$\Delta=4f^3+27g^2$. We emphasize that for  $T$ being an abstract
variable, we do not obtain a vanishing of $\Delta$. However, we see
that 
\beq
\left.(4f^3+27g^2)\right\vert_{T=0}\sim s_4 s_8^3 - s_3 s_8^2 s_9+ s_2 s_8 s_9^2 - s_1 s_9^3\,,
\eeq
which agrees precisely with $t$ given in \eqref{eq:SU2divisor}.
Thus, for the special choice  $T\equiv t$ we obtain a vanishing of $\Delta$ to first order. In fact, if we set 
$T\equiv t$ we see that $\Delta$ vanishes also to second order at $t=0$ due to additional cancellations. We then obtain
\beq \label{eq:Delta}
	\Delta=t^2\Delta'\,,\qquad \Delta'= 4 s_1 s_3^3 + 4 s_2^3 s_4 - 18 s_1 s_2 s_3 s_4 + 27 s_1^2 s_4^2-s_2^2 s_3^2 \,.
\eeq
Here, the remainder $\Delta'$ of the discriminant is in the class $[\Delta']=-6K_B+2S_7-4S_9$ so that $[\Delta]=[\Delta']+2[t]=-12K_B$.

In summary, we see that the singularity structure of the elliptic
fibration defined by the Weierstrass model with \eqref{eq:WSFSU2}
crucially depends on the particular form of $t=0$ with triple point
singularities at $s_8=s_9=0$. In particular, the forms in
\eqref{eq:WSFSU2} do not have the structure needed for an SU(2)
singularity through Tate's algorithm
\cite{tate1975algorithm,Bershadsky:1996nh}, and do not have the form
expected for an SU(2) on a smooth divisor
$t=0$, because the induced ring of local functions is not a universal
factorization domain  \cite{Morrison:2011mb}. Thus, we refer to
the model \eqref{eq:WSFSU2} and models of similar type more generally
as \textit{non-Tate form} Weierstrass models. Explicitly, we observe
that the Tate coefficients
\bea \label{eq:SU2Tate}
& a_1=a_3=0\,,\qquad a_2=-s_3 s_8 - s_2 s_9\,,\qquad a_4=s_2 s_4 s_8^2 +(s_2 s_3  - 3 s_1 s_4) s_8 s_9 + s_1 s_3 s_9^2&\nn\\
 &a_6=-s_1 s_4^2 s_8^3+(2 s_1 s_3 -s_2^2  )s_4 s_8^2s_9+(2 s_2 s_4- s_3^2 )s_1  s_8s_9^2-s_1^2 s_4 s_9^3&
\eea
for \eqref{eq:WSFSU2} that naively follow from \eqref{eq:Tate} by the
tuning \eqref{eq:simpleSol} do not exhibit the vanishing orders in
Tate's algorithm for the realization of an SU(2) gauge group
\cite{tate1975algorithm,Bershadsky:1996nh}.

We conclude by noting that \eqref{eq:WSFSU2} assumes the
normal form of a Weierstrass model with $I_2$ singularities as dictated by Tate's algorithm if 
$s_8$ or $s_9$ are constants, {\it i.e.}~$t=0$ is smooth. For example,
if $s_9=
{\rm const.}$ we 
can shift the variables so that $t\equiv s_1$ and \eqref{eq:WSFSU2} assumes the form of a 
Weierstrass model with $I_2$ singularities in  \cite{Morrison:2011mb}.

\subsection{The non-Abelian matter spectrum}
\label{sec:NonAbelMatter}

We are now in a position to determine the matter spectrum of F-theory on
the Calabi-Yau manifold $X^{\text{SU(2)}}$.  For the reader only
interested in the results of this analysis, we summarize the matter
content in Table \ref{tab:XSU2matter}.
\begin{table}[ht!]
\begin{center}
\footnotesize
\renewcommand{\arraystretch}{1.2}
\begin{tabular}{|c|c|c|@{}c@{}|}
\hline
 SU(2)-rep & Multiplicity & Fiber & Locus  \\ \hline
$\mathbf{4}$ & $x_{\mathbf{4}}=\frac12\cS_9\cdot (-K_B+\cS_9-\cS_7)$ & $I_0^{*ns}$  & $V_{\text{Sing}}=\{s_8 = s_9 = 0\}$ \rule{0cm}{.4cm}  \\[.1cm] \hline

$\mathbf{3}$ &  $x_{\mathbf{3}}=\frac12[t]\cdot([t]+K_B) + 1 - 6
 x_{\bf 4}$ & $I_2$
 &  $V_{\text{SU}(2)}=\{t=0\}$ \rule{0cm}{.4cm}
 \\[0.1cm] \hline
$\mathbf{2}$  &\rule{0cm}{0.7cm}
$\begin{array}{c}
 x_{\mathbf{2}}=2 (3 K_B^2-K_B\cdot(2 \cS_7-\cS_9) \\
 - \cS_7^2 + \cS_7\cdot \cS_9 - \cS_9^2)+2 x_{\mathbf{4}}
\end{array}  $ & $I_3$ & $
V(\mathfrak{p}_1)\cup V_{\text{Sing}}$\\[0.1cm] \hline
\end{tabular}
\caption{\label{tab:XSU2matter} Matter spectrum of $X^{\text{SU(2)}}$. Shown is the multiplicity of full hypermultiplets in a 6D SUGRA theory. We note that there is only a 
half-hypermultiplet in the $\mathbf{4}\oplus \mathbf{2}\oplus\mathbf{2}$ at each ordinary triple point $s_8=s_9=0$ of $t=0$.}
\end{center}
\end{table}

We begin with the matter content localized at codimension one.  As
noted before, the SU(2) gauge algebra is supported on a Riemann
surface $t=0$ of higher (arithmetic) genus $g$, which is computed via
\eqref{eq:arithmeticGgeneral}. As $t=0$ has a number of 
$[s_8]\cdot [s_9]$ ordinary triple point singularities, each of which
contribute $3$ to $g$, we obtain the topological genus $p_g$
\beq
	p_g=g-3[s_8]\cdot [s_9]\,,
\eeq
which is explicitly given in the first line of
\eqref{eq:geometricGenus}.  
In a 6D compactification, the topological
genus $p_g$ 
gives rise to $p_g$ hypermultiplets in the
adjoint representation $\mathbf{3}$ of the SU(2) gauge group on $t=0$
\cite{Witten:1996qb}.  
Employing \eqref{eq:cubicsections}, this gives
 the multiplicity $x_{\mathbf{3}}$ in the second row in Table
\ref{tab:XSU2matter}.

Next let us consider the matter contribution of the triple point
singularities at the loci $s_8 = s_9 = 0$.  One way to attain a triple
point singularity on a divisor supporting an SU(2) is to take a Tate
model for an SU(2) on a smooth divisor  $\tilde{t}$, and then to deform
the divisor to get a triple point singularity.  In this scenario, the
triple point can be viewed as a limit of three double point
singularities.  Furthermore, each double point is reached in a limit
of a family of smooth surfaces; reasoning following
\cite{Anderson:2015cqy}, each such double point must be associated
with an adjoint representation since there is no intermediate
opportunity for a matter transition through a superconformal fixed
point, and for similar reasons the triple point in the Tate
construction must then represent three adjoint matter multiplets.  For the
non-Tate model found here, however, the arithmetic
genus three singularity may give a matter
content with a half-hypermultiplet in the triple symmetric {\bf 4}
representation.  To distinguish these possibilities, further analysis
is needed.  In the following section we argue that by matching the
matter content with the Higgsed U(1) theory, the only consistent
possibility is that each triple point carries a half-hypermultiplet in
the {\bf 4} representation.  This gives the 
multiplicity $x_{\mathbf{4}}$ in the first row in Table
\ref{tab:XSU2matter}.

Another approach, in principle, to determining the matter content at
the intersection point is to explicitly resolve the singularity of the
Calabi-Yau manifold over the triple intersection point.  This is an
interesting direction for study, which we leave the details of for
future work.  We make several comments, however.  First, the local
analysis will determine the representation of SU(2)$\times$SU(2)$\times$ SU(2) 
realized at the intersection of three independent
divisors.  This will either give three bifundamental type
representations, corresponding to the possibility of three adjoints
for the SU(2) on the connected divisor, or a
trifundamental\footnote{The possibility of a trifundamental
  representation arising at a triple point of an $I_2$ locus was also
  discussed in \cite{Bhardwaj:2015oru}.}
representation {\bf 2}$\times${\bf 2}$\times${\bf 2}, which would
break up into a {\bf 4} and two fundamental {\bf 2}'s when the divisor
is connected and we embed $SU(2) \subset SU(2) \times SU(2) \times
SU(2)$.  (Actually, we would get a half trifundamental, as this
representation is self-conjugate).  
Note that while for a larger group like SU(3), the precise matter
content, such as the presence of an adjoint  {\it vs.} a symmetric +
antisymmetric, depends on how the divisor connects to itself, 
{\it i.e.} on whether the local representation on each branch is
fundamental or antifundamental, that distinction is irrelevant for
SU(2) where the fundamental representation is self-conjugate.
In any case, this analysis suggests that when the triple point gives a
triple-symmetric {\bf 4} representation there will also be two
fundamental {\bf 2} representations present.

The Kodaira singularity at the
triple points is of type $I_0^{*}$. 
Since this is a codimension two singularity, the split/non-split
distinction and monodromy structure
is not relevant in the same way as it is for codimension
one singularities, where it would determine whether the gauge group would
be $G_2$ or SO(8).
For six-dimensional theories, this singularity arises at a point, so
there is no question of monodromy, and the Dynkin diagram associated
with the singularity is a $D_4$.  
Locally,
the matter structure associated with the codimension two singularity
is determined by the embedding of the three
single nodes associated with the $A_1$ SU(2) factors on the
branches of the $I_2$ locus into the $D_4$.
This can be done in  an essentially unique way that respects the
permutation symmetry on the $A_1$ factors
by embedding the three  $A_1$ factors as the three outer nodes
of the Dynkin diagram $D_4$. The central node
then represents a matter state that is charged under all three SU(2)
factors, and thus associated with
the trifundamental representation 
$\mathbf{2}\times\mathbf{2}\times\mathbf{2}$, which yields
the $\mathbf{4} +\mathbf{2}+\mathbf{2}$ representation upon the
embedding of  SU$(2)\subset\text{SU}(2)\times\text{SU}(2)\times\text{SU}(2)$ by identifying 
the three SU(2) factors as discussed above.
This gives strong evidence from the group theory point of view
that indeed the local $D_4$ structure at the triple point must be
associated with the ${\bf 4}$ representation of the SU(2) on the $I_2$
locus.  A more explicit resolution of this singularity is left to
future work.
Note that for 4D F-theory models, the codimension two $D_4$ singularity
arises over a curve in the base threefold.  While there may be
nontrivial monodromy around this curve, this simply corresponds to the
identification of the different SU(2) factors on the branches of the
$I_2$ locus
that enter the triple point.  Since these branches are already
identified globally, this does not modify the above conclusion that
the resulting matter content should include the {\bf 4} representation
of the SU(2).

Finally, we use the Weierstrass model \eqref{eq:WSFSU2} to find the
codimension two singularities of $X^{\text{SU}(2)}$ at the
intersection $t=\Delta'=0$ with $\Delta'$ given in
\eqref{eq:Delta}. The computation of the primary decomposition of the
ideal $I:=\{t,\Delta'\}$ yields two prime ideals, which we denote by
$\mathfrak{p}_1$ and $\mathfrak{p}_2$. As these ideals are generated
by 14 and six polynomials, respectively, we do not present their
explicit forms here. Consequently, the variety $V(I)$ is reducible
with irreducible components $V(\mathfrak{p}_1)$ and
$V(\mathfrak{p}_2)$ that can be shown, employing the resultant
technique as in \cite{Cvetic:2013nia}, to have multiplicities $1$ and
$2$ inside $V(I)$, respectively. Thus, we find the homology relation
\beq
 [V(I)]=[V(\mathfrak{p}_1)]+2[V(\mathfrak{p}_2)]\,.  
\eeq
The
individual homology classes are computed as explained in
\cite{Cvetic:2013uta} to be
\begin{eqnarray} \label{eq:Multiplicitiesp1p2}
 [V(\mathfrak{p}_1)]  & = & 2 (3 K_B^2 -K_B\cdot(2\cS_7-\cS_9) - \cS_7^2  + \cS_7\cdot \cS_9 - \cS_9^2) \,,  \nonumber \\
 ~[V(\mathfrak{p}_2)] & = & 6 K_B^2 - K_B\cdot ( \cS_9-2\cS_7) + 3 \cS_7\cdot \cS_9 - 3 \cS_9^2\,.
\end{eqnarray}
Next, we determine the singularity type of $X^{\text{SU}(2)}$ along
these two irreducible components. By reducing the Weierstrass
coefficients $f$, $g$ and the discriminant $\Delta$ given in
\eqref{eq:WSFSU2} and \eqref{eq:Delta} as well as the Tate
coefficients \eqref{eq:SU2Tate} modulo the ideals $\mathfrak{p}_1$,
$\mathfrak{p}_2$, respectively, we find Kodaira singularities of type
$I_3$ and $III$, respectively. Thus, the locus $V(\mathfrak{p}_1)$
supports a number of $[V(\mathfrak{p}_1)]$ matter fields in the
fundamental representation $\mathbf{2}$ of SU(2), as shown in the last
line of Table \ref{tab:XSU2matter}, while no matter fields are located
on $V(\mathfrak{p}_2)$ since the type $III$ fiber is just a
degenerated $I_2$ fiber with no additional $\mathbb{P}^1$ harboring
matter states.  In a compactification on a threefold
$X^{\text{SU}(2)}$ to 6D, the found matter fields form a full
hypermultiplet. The multiplicity of matter fields in the $\mathbf{2}$
representation is given in the last line of Table
\ref{tab:XSU2matter}, where we have added $[s_8]\cdot[s_9]$
fundamentals contributed by the ordinary triple point singularities of
$t=0$, matching the analysis of
the local trifundamental representation mentioned above.

We conclude by noting that the anomaly coefficient $b$ of the 6D SUGRA theory given by 
F-theory on the threefold $X^{\text{SU(2)}}$ is given by the class of $t$, {\it i.e.}~it reads
\beq\label{eq:bSU2}
b^{\text{SU}(2)}=[t]=	-2K_B+[s_8]+[s_9]=-3K_B+2\cS_9-\cS_7\,.
\eeq
Employing this coefficient, the spectrum in Table
\ref{tab:XSU2matter}, $a=K_B$, and the anomaly coefficients
$(A_\mathbf{R},B_\mathbf{R},C_\mathbf{R})=(1,0,\frac12),\,(4,0,8),\,(10,0,41)$
for the SU(2)-representations
$\mathbf{R}=\mathbf{2},\,\mathbf{3},\,\mathbf{4}$, respectively, we
readily check that the two 6D gauge and mixed gauge-gravity anomalies
are cancelled. 
For the anomaly cancellation to work, following the genus analysis, it
is necessary that there is only a
half-hypermultiplet in the representation $\mathbf{4}\oplus
\mathbf{2}\oplus\mathbf{2}$ at each triple point singularity of $t=0$,
as indicated in Table \ref{tab:XSU2matter}. 
Note furthermore, as mentioned earlier,
that there is an anomaly equivalence
\begin{equation}
\frac{1}{2}{\bf 4} + 7 \times {\bf 2} \leftrightarrow
3 \times {\bf 3} + 7 \times {\bf 1}.
\label{eq:anomaly-equivalence}
\end{equation}
This shows that with the number of matter fields in the fundamental
identified above, it is not possible to satisfy the anomaly conditions
when the triple intersection point supports three adjoints and any
positive number of fundamental representations.  This provides an
alternative argument using only anomaly conditions and counting of
known singularity types that the matter content at the triple points is
$\frac{1}{2}\times {\bf 4} + {\bf 2}$ as identified above.  For more
details on the relevant anomaly cancellation conditions in the context of
F-theory, see {\it e.g.}~the review \cite{Taylor:2011wt}.
Finally, note that as found in \cite{Anderson:2015cqy}, we expect that
the total number of fields that must be brought together to explicitly
undergo a transition like (\ref{eq:anomaly-equivalence}) will
bring the theory to a superconformal transition point, where a tensor
branch is also available.  A more explicit treatment of such
transitions will be presented elsewhere.

\subsection{Matching effective theories through the Higgs transition}
\label{sec:MatchingHiggsing}

Next, we match the effective field theory of F-theory on
$X^{\text{SU}(2)}$ with the Abelian model obtained by F-theory on $X$.
We show that the two theories are related under a Higgsing by matter
in the adjoint representation.  As mentioned above, this corresponds to
the extremal transition $X^{\text{SU}(2)}\rightarrow X$ induced by
switching on the deformation parameter $z_1$ defined in
\eqref{eq:unHiggs}.

We begin by matching the charged matter spectrum of the non-Abelian model in Table \ref{tab:XSU2matter} with the one of the 
Abelian model in Table \ref{tab:poly3_matter} through the adjoint Higgsing. 
First, we note the following branching of SU(2)
representations under the breaking $\text{SU}(2)\rightarrow \text{U}(1)$:
\bea
	\mathbf{4}\rightarrow \mathbf{1}_{3}\oplus  \mathbf{1}_{-3}\oplus \mathbf{1}_{1}\oplus \mathbf{1}_{-1}\,,\qquad \mathbf{3}\rightarrow \mathbf{1}_{2}\oplus \mathbf{1}_{-2}\oplus\mathbf{1}_0\,,\qquad \mathbf{2}\rightarrow \mathbf{1}_1\oplus \mathbf{1}_{-1}\,.
\eea
Here we have computed U(1)-charges using the generator $2\sigma_3$, where $\sigma_3$ is the third Pauli matrix of SU(2). 
Next, we use the fact that a hypermultiplet with charge $q$ is composed of states with charge $q$ and $-q$ to 
eliminate negative charges. Finally, employing that
two hypermultiplets with charges $q=\pm 2$, respectively, from the adjoint representation are eaten up in the Higgsing by the 
massive W-bosons of the broken SU(2) vector multiplet, we obtain an Abelian theory
with the following numbers $x_{\mathbf{1}_q}$ of hypermultiplets with charges $q=1,2,3$:
\bea
	x_{\mathbf{1}_3}=2x_{\mathbf{4}}\,,\qquad x_{\mathbf{1}_2}=2 (x_{\mathbf{1}_3} - 1)\,,\qquad x_{\mathbf{1}_1}=2(x_{\mathbf{1}_4}+x_{\mathbf{1}_1})\,.
\eea
Comparing with the matter spectrum in Table \ref{tab:poly3_matter},
using Table \ref{tab:XSU2matter}, we see that we precisely reproduce
the effective theory of F-theory on $X$. 
Furthermore, we note that the
anomaly coefficient $b$ in \eqref{eq:heightF3} of the Abelian theory
is $2b^{\text{SU}(2)}$ with $b^{\text{SU}(2)}$ given in
\eqref{eq:bSU2} as expected.  This in particular implies an anomaly
free theory in 6D.
This precise matching between the spectra gives a rigorous argument for
the presence of {\bf 4} matter at the triple point singularities,
matching with the results of the arguments given in the previous section; this
is the only matter content that would give a consistent U(1) theory
after Higgsing.

Next, we note that the number of complex structure moduli increases in
the Higgsing, corresponding geometrically to the deformations
$X^{\text{SU}(2)}\rightarrow X$. The new complex structure moduli are
naturally associated with the deformation parameters in $z_1$. 
We expect therefore that the number of independent parameters that
deform
$z_1$ away from the locus $z_1 = 0$
will match
the number of Higgs VEVs, {\it i.e.}~neutral hypermultiplets in the
$\mathbf{3}$ representation.\footnote{Note that there is no D-term
  condition in an adjoint Higgsing.} 
As there are $x_{\mathbf{3}}=p_g$
matter fields in the $\mathbf{3}$ representation, each of which has
one neutral component, we expect $p_g$ new moduli and deformation
parameters in $z_1$.  
To be concrete,
for the concrete base $B=\mathbb{P}^2$ we can
compute the change in the number of complex structure moduli by a
counting of monomials in appropriate classes.  First, we compute the
number $x_{\mathbf{3}}$ of adjoint fields in the representation
$\mathbf{3}$ and Higgs VEVs according to Table \ref{tab:XSU2matter} as
\beq \label{eq:x3P2}
	x_{\mathbf{3}}=28 - \tfrac{15}2 \cS_7 + \tfrac12\cS_7^2 + 6 \cS_9 +  \cS_7\cdot \cS_9 - \cS_9^2\,,
\eeq
where we have used that $K_{B}=\mathcal{O}_{\mathbf{P}^2}(-3)$.
Explicitly computing the number of deformation parameters in $z_1$,
assuming the generic form  (\ref{eq:UFDSolution}) for the solution to
$z_1 = 0$,  we can parameterize the deformations by replacing
$\sigma_5, \sigma_7$ by generic $s_5, s_6, s_7$.  The number of
independent monomials in a degree $d$ divisor class is $m[d] =
(d +1) (d +
2)/2$, allowing us to confirm that the number of independent degrees
of freedom that deform $z_1 \neq 0$ is
\begin{equation}
m[s_5]+m[s_6]+m[s_7]-m[\sigma_5]-m[\sigma_7]  = x_{\bf 3}\,.
\label{eq:defsinz1}
\end{equation}
In principle, it should also be possible to check whether the number of
independent Weierstrass moduli in both the SU(2) and U(1) models
involved match precisely with the number of neutral scalar fields
expected from the gravitational anomaly cancellation condition $H-V=
273-29T$.  While the computation just performed demonstrates that the
difference between these numbers is correctly captured by the
deformation parameters in $z_1$, there is some redundancy in our
parameterization of these models through the $s_i$'s; removing this
redundancy and identifying the proper number of independent degrees of
freedom in the Weierstrass model would a useful check to determine
whether the models presented here are the most general forms for the
given spectra, or only represent a subset of the possibilities.

\subsection{Models over $B=\mathbb{P}^2$}
\label{sec:NonAbelianModlesP2}

We conclude the discussion of F-theory compactified on the Calabi-Yau
manifold $X^{\text{SU}(2)}$ with the concrete models obtained for
$B=\mathbb{P}^2$.

We begin by considering the generic class of SU(2) models on $\P^2$.
When the SU(2) is realized on a smooth divisor of degree $d$, the
genus of the corresponding curve is $g = (d -1) (d - 2)/2$.  This is
the number of matter fields in the adjoint ({\bf 3}) representation.
From explicit construction or anomaly cancellation, it is
straightforward to determine that the number of fundamental ({\bf 2})
matter fields is $x_{\bf 2} = 16 + 6d^2 -16g$.  This parameterizes the
full spectrum of F-theory constructions on $\P^2$ with an SU(2) gauge
group realized on a smooth divisor.  Using the anomaly equivalence
(\ref{eq:anomaly-equivalence}), we expect that we can exchange 3
adjoints and seven uncharged moduli in any of these models for a
half-hypermultiplet in the {\bf 4} representation and seven
fundamentals.  For example, when $d = 8$, we have a genus 21 curve,
and the generic matter content consists of 21 adjoints and 64
fundamentals.  We would expect anomaly-equivalent models with $21-3x$
adjoints, $x$ half-hypermultiplets in the {\bf 4} representation, and
$64 + 7 x$ hypermultiplets in the fundamental representation.  These
classes of models
(for general $d$) comprise all models that are consistent from the
low-energy 6D supergravity point of view, and that have no tensor
multiplets, an SU(2) gauge group, and matter in only the {\bf 1, 2,
  3, 4} representations.  We might hope to identify in F-theory using
the approach described here all such models that have at least one
adjoint representation that can be Higgsed to give a U(1) theory with
charges up to $q = 3$.

Next we recall that the Calabi-Yau manifold $X^{\text{SU}(2)}$ is
defined by \eqref{eq:PF3} with tuned complex structure so that
$s_5\equiv s_6\equiv s_7\equiv 0$. Thus, the model exists as long as
all other sections $s_i$ exist, {\it i.e.}~are associated to effective
divisor classes. By explicitly solving the effectiveness conditions
implied by this, we again obtain the allowed region in Figure
\ref{fig:allowedregion}. For every Abelian model $X$ there exists a
corresponding model $X^{\text{SU}(2)}$ and vice versa. For each of
these 16 inequivalent models (recall the $\mathbb{Z}_2$-symmetry
\eqref{eq:Z2symmetry}) we readily compute all divisor classes $[s_i]$,
the class of the SU(2)-divisor $t=0$ as well as the charged matter
spectrum in Table \ref{tab:XSU2matter}. We obtain:
\beq\label{eq:NonAbelspectrumP2} \text{
\begin{tabular}{c||c|c|c|c|c|c|c||c|c}
$(\cS_7,\cS_9)$ &$[s_1]$&$[s_2]$&$[s_3]$&$[s_4]$&$[s_5]$&$[s_6]$&$[s_8]$& $[t]$ & $(x_{\mathbf{4}},x_{\mathbf{3}},x_{\mathbf{2}})$\\
\hline
	$(0,0)$&9&6&3&0&6&3&3&9&$(0,28,54)$\rule{0pt}{13pt} \\
	$(1,0)$&8&6&4&2&5&3&2&8&$(0,21,64)$\rule{0pt}{12pt} \\
	$(2,0)$&7&6&5&4&4&3&1&7&$(0,15,70)$\rule{0pt}{12pt} \\
	$(3,0)$&6&6&6&6&3&3&0&6&$(0,10,72)$\rule{0pt}{12pt} \\
	$(1,1)$&7&5&3&1&5&3&3&10&$(\tfrac32,27,61)$\rule{0pt}{12pt} \\
	$(2,1)$&6&5&4&3&4&3&2&9&$(1,22,68)$\rule{0pt}{12pt} \\
	$(3,1)$&5&5&5&5&3&3&1&8&$(\tfrac12,18,71)$\rule{0pt}{12pt} \\
	$(1,2)$&6&4&2&0&5&3&4&12&$(4,31,56)$\rule{0pt}{12pt} \\
	$(2,2)$&5&4&3&2&4&3&3&11&$(3,27,64)$ \rule{0pt}{12pt} \\
    $(3,2)$&4&4&4&4&3&3&2&16&$(2,24,68)$ \rule{0pt}{12pt} \\
	$(2,3)$&4&3&2&1&4&3&4&13&$(6,30,58)$ \rule{0pt}{12pt} \\
	$(3,3)$&3&3&3&3&3&3&3&12&$(\tfrac92,28,63)$ \rule{0pt}{12pt} \\
	$(2,4)$&3&2&1&0&4&3&5&15&$(10,31,50)$ \rule{0pt}{12pt} \\
	$(3,4)$&2&2&2&2&3&3&4&14&$(8,30,56)$ \rule{0pt}{12pt} \\
	$(3,5)$&1&1&1&4&3&3&5&16&$(\tfrac{25}{2},30,47)$ \rule{0pt}{12pt} \\
	$(3,6)$&0&0&0&0&3&3&6&18&$(18,28,36)$ \rule{0pt}{12pt} \\
\end{tabular}
}
\eeq

There are some remarks in order. First, we note that in the absence of
triple point singularities of $t=0$, its minimal degree is
$6$. However, in that case the model $X^{\text{SU}(2)}$ is completely
equivalent to the elliptic fibrations by quartics in
$\text{Bl}_1\mathbb{P}^2(1,1,2)$ of Morrison, Park
\cite{Morrison:2012ei}, as mentioned before. Thus, there have to
exist models with $[t]=3,4,5$.
As discussed before at the end of
Section \ref{sec:GeoUnHiggsing}, these can be obtained from $X$ if we
relax the effectiveness condition on $[s_8]$. Indeed, we
can then lower the degree
of $[t]=[s_1]$ to $3$, as expected. 

Second, in the case with ordinary triple point singularities on $t$,
we observe that our list \eqref{eq:NonAbelspectrumP2} does not produce
all models that seem geometrically possible. For example, a model with
$[t]=5$ has an arithmetic genus of $g=6$ which seems to allow for one
ordinary triple point singularity while still exhibiting a geometric
genus $p_g=3$, {\it i.e.}~adjoints for a Higgsing to an Abelian
theory. Similar models with a different number of ordinary triple
points than in \eqref{eq:NonAbelspectrumP2} seem to be constructable
also for higher degree curves $t=0$.  
Naively it would seem that we can simply choose, for example $[s_1]
=[s_2] =[s_3] =[s_4] = 2$ and $[s_8] =[s_9] = 1$ in the Weierstrass
form
(\ref{eq:WSFSU2}).  While this set of choices are not compatible with
effectiveness of all divisor classes in (\ref{eq:cubicsections}), this would seem to
define a well-defined Weierstrass model with the SU(2) structure of
interest realized on a quintic curve with a single triple point at the
intersection $s_8 = s_9 = 0$.  The issue, however, is that since $f$
is of degree 12 and $g$ of degree 18 in homogeneous coordinates
$[x:y:z]$, this leads to a problematic (6, 12) singularity when $z
\rightarrow 0$.  The compatibility of the divisor classes with
(\ref{eq:cubicsections}) avoids this problem.
It would be interesting to
understand whether the absence of these models is a mere artifact of
how the Weierstrass form \eqref{eq:WSFSU2} is constructed, or whether
this is an indication of a fundamental limitation in the spectrum of
models available from F-theory, or even in 6D supergravity consistent
with quantum gravity constraints.  A systematic mathematical
classification of Weierstrass models of elliptic fibrations with $I_2$
singularities over singular divisors would help to answer this
question.

\section{Further unHiggsing to larger non-Abelian groups}
\label{sec:further-unHiggsing}

In this section we discuss the possibility to further unHiggs the
non-Abelian model defined by F-theory on $X^{\text{SU}(2)}$.  Here, we
are motivated by the search for a resulting non-Abelian theory that
has a standard matter spectrum consisting only of fundamentals,
anti-fundamentals and adjoints. In this case, the geometric
realization of the corresponding elliptic fibration should follow the
standard rules of Tate's algorithm. Starting with these standard Tate
Weierstrass models the inverse process of the unHiggsing described here
can then be understood as a deformation (re-Higgsing) of these
Weierstrass models to a non-Tate Weierstrass model. Systematizing this
deformation procedure outlined below may shed light on the general
construction of non-Tate Weierstrass forms with novel matter
structures in F-theory.  For a recent application of this idea, we
refer the reader to \cite{Anderson:2015cqy}.

Here, we discuss two unHiggsing, one to models with  $G_2\times \text{SU}(2)$ gauge group 
and standard matter content given by adjoints and (bi-)fundamentals
and one to models with $\text{SU}(2)\times \text{SU}(2)\times
\text{SU}(2)$ gauge group and with a matter content
that includes trifundamental matter.

\subsection{UnHiggsing SU(2) with the $\mathbf{4}$ representation to $\text{SU}(2)\times G_2$}

One possible unHiggsing of F-theory on $X^{\text{SU}(2)}$ yields a theory with 
$G_2\times\text{SU}(2)$ gauge group on two different divisors and with a standard matter 
spectrum consisting of adjoints, fundamentals and bifundamentals. The unHiggsing is achieved 
by imposing
\beq \label{eq:G2tuning}
	s_8\equiv a s_9\,
\eeq
for an appropriate section $a\in \mathcal{O}(-K_B-\mathcal{S}_7)$, which can exist if $-K_B-\mathcal{S}_7$ is an effective class (if $[s_9]\geq [s_8]$, we can impose the inverse relation 
$s_9=bs_8$ for appropriate $b$.).

With this tuning,  the SU(2) divisor $t=0$ defined in
\eqref{eq:SU2divisor}  degenerates as
\beq
 t=s_9^3(s_4 a^3- s_3 a^2 + s_2 a - s_1 )\,,
\eeq
so that its triple point singularities disappear at the cost of an overall factor of $s_9^3$.
Indeed, the Weierstrass model \eqref{eq:WSFSU2} reduces to the form
\bea \label{eq:WSFG2SU2}
	&f=(-\tfrac13 \tilde{s}_2^2 +  \tilde{s}_3 \tilde{s}_1) s_9^2\,,\qquad  g=(-\tfrac{2}{27}  \tilde{s}_2^3 +\tfrac13 \tilde{s}_2 \tilde{s}_3\tilde{s}_1 -  s_4\tilde{s}_1^2) s_9^3\,,&\nn\\
&\Delta=-16 \tilde{s}_1^2 \tilde{s}_9^6\Delta'\,,\qquad \Delta'=-\tilde{s}_2^2 \tilde{s}_3^2+4 \tilde{s}_1 \tilde{s}_3^3+4 \tilde{s}_2^3 \tilde{s}_4-18 \tilde{s}_1 \tilde{s}_2 \tilde{s}_3 \tilde{s}_4+27 \tilde{s}_1^2 \tilde{s}_4^2\,,&
\eea
where we made the definitions
\beq \label{eq:stilde}
	\tilde{s}_1=s_1+a s_2 - a^2 s_3 + a^3 s_4\,,\qquad\tilde{s}_2=s_2-2 a s_3 + 3 a^2 s_4\,,\qquad \tilde{s}_3=s_3-3 a s_4\,.
\eeq
The Weierstrass form \eqref{eq:WSFG2SU2} reveals the presence of singularities of Kodaira types $I_2$ at $\tilde{s}_1=0$ and  
$I_0^*$ at $s_9=0$, respectively. We readily observe that \eqref{eq:WSFG2SU2} is of the normal form of a Weierstrass model 
with $I_2$ singularity following from Tate's algorithm or the analysis in \cite{Morrison:2011mb}.
Using the orders of vanishing of the Tate coefficients \eqref{eq:Tate} in the limit 
\eqref{eq:G2tuning}, which are $(\infty,1,\infty,2,3)$, or by computing the irreducible 
monodromy cover \cite{Grassi:2011hq}, we see that the singularity at $s_9=0$ is non-split, 
{\it i.e.}~of type $I_0^{*\text{ns}}$ yielding a $G_2$ gauge symmetry 
\cite{Bershadsky:1996nh}. Thus, F-theory on $X^{\text{SU}(2)}$ with the tuning \eqref{eq:G2tuning}  has the gauge group
\beq
	G=\text{SU}(2)\times G_2\,.
\eeq

 Note that the Weierstrass form \eqref{eq:WSFG2SU2}, like
(\ref{eq:WSFSU2}), are acceptable for choices of $s_1,
s_9$ that violate the effectiveness conditions
(\ref{eq:cubicsections}). However, if in addition also \eqref{eq:stilde} is to be satisfied, \textit{i.e.}~if the model shall be deformable 
back to $X^{\text{SU}(2)}$, such models suffer from the same issue
discussed earlier  and have problems with bad singularities at
infinity.  
For example, there should be no problem in tuning, for
example, a $G_2$ on a line $[s_9] = 1$ and an SU(2) on a conic 
$[s_1]= 2$.  This, however, would imply that
 $[s_8]=-2$, {\it i.e.}, that \eqref{eq:stilde} breaks down.
As we see below, in this case there is insufficient matter to carry
out the Higgsing that is needed to
deform the model to return to the SU(2) models where \eqref{eq:WSFSU2} is valid,
explaining the absence of a corresponding SU(2) model.

The matter content of the F-theory effective field theory can be derived from the Weierstrass 
model \eqref{eq:WSFG2SU2}. As we will discuss, due to the presence of the $G_2$ gauge group,
matter representations arise both at codimension one, \textit{i.e.}~are non-local, as well as at 
codimension two loci where the singularities of the elliptic fibration enhance. Before presenting
the details of this analysis, we summarize the derived matter spectrum in Table \ref{tab:XSU2G2matter}. We emphasize again that the spectrum only contains fundamental and adjoint representations, which can be  attributed to 
the smoothness of both gauge divisors $\tilde{s}_1=0$ and $s_9=0$ as well as the standard form of the Weierstrass model. 
We note that there is an additional Kodaira singularity of type $III$ at
the codimension two locus $\tilde{s}_1=\tilde{s}_2=0$ that does not give rise to matter fields.
\begin{table}[ht!]
\begin{center}
\footnotesize
\renewcommand{\arraystretch}{1.2}
\begin{tabular}{|c|c|c|@{}c@{}|}
\hline
 Rep \!\!&\!\! Multiplicity \!\!&\!\! Fiber & Locus  \\ \hline
$(\mathbf{2},\mathbf{7})$ \!\!&\!\! $
x_{(\mathbf{2},\mathbf{7})}=\tfrac12\cS_9\cdot (-3K_B-\cS_7-\cS_9)$ \!\!&\!\! $I_2^{*}$  & $V_{\text{bf}}=\{\tilde{s}_1 = s_9 = 0\}$ \rule{0cm}{.4cm}  \\[.1cm] \hline

$(\mathbf{1},\mathbf{7})$ \!\!&\!\!  $x_{(\mathbf{1},\mathbf{7})}=\cS_9\cdot (-2K_B+\cS_7-\cS_9)$ \!\!&\!\! $-$
 &  non-local \rule{0cm}{.4cm}
 \\[0.1cm] \hline
$(\mathbf{2},\mathbf{1})$ \!\!&\!\!\rule{0cm}{0.4cm}
$ x_{(\mathbf{2},\mathbf{1})}=\tfrac{1}{2} (-4 K_B + 4 \mathcal{S}_7 - 3 \mathcal{S}_9) (-3 K_B - \mathcal{S}_7 -\mathcal{S}_9)  $ \!\!&\!\! $I_3$ & $
\begin{array}{rl}
V_2\!\!\!&\!\!=\{\tilde{s}_1=4 \tilde{s}_2 \tilde{s}_4-\tilde{s}_3^2=0\}\\\!\!&\!\!\cup V_{\text{bf}}
\end{array} $\\[0.1cm] \hline
$(\mathbf{3},\mathbf{1})$ \!\!&\!\! $ x_{(\mathbf{3},\mathbf{1})}=1+\tfrac12(-3K_B-\cS_7-\cS_9)\cdot (-2K_B-\cS_7-\cS_9)$ \!\!&\!\! $I_2$ &$V_{\text{SU}(2)}=\{\tilde{s}_1=0\}$\rule{0cm}{.4cm}  \\[0.1cm]  \hline
$(\mathbf{1},\mathbf{14})$ \!\!&\!\! $x_{(\mathbf{1},\mathbf{14})}=1+\tfrac12\cS_9\cdot(\cS_9+K_B)$ \!\!&\!\! $I_0^{*}$ & $V_{\text{G}_2}=\{s_9=0\}$ \rule{0cm}{.4cm} \\[0.1cm] \hline 
\end{tabular}
\caption{\label{tab:XSU2G2matter} Matter spectrum of F-theory on
  $X^{\text{SU(2)}}$ with the tuning \eqref{eq:G2tuning} to a model
  with gauge group $\text{G}_2\times \text{SU}(2)$. Shown are the
  multiplicities of full hypermultiplets in a 6D SUGRA theory.} 
\end{center}
\end{table}

Cancellation of 6D anomalies can be checked using the group theory
coefficients $(A_\mathbf{R},B_\mathbf{R},C_\mathbf{R})=(1,0,\tfrac14),
(4,0,\tfrac52),(1,0\tfrac12),(4,0,8)$ for the $G_2$-representations
$\mathbf{R}=\mathbf{7},\mathbf{14}$ and the SU(2)-representations
$\mathbf{R}=\mathbf{2},\mathbf{3}$, respectively, given for example in
\cite{Erler:1993zy}. The coefficients
$b^{\text{SU}(2)}=[s_1]=-3K_B-\mathcal{S}_7-\mathcal{S}_9$ and
$b^{G_2}=\mathcal{S}_9$ enter the 6D GS-counterterms. 

Next, we explain the derivation of the matter spectrum given in Table \ref{tab:XSU2G2matter}.
We begin with the non-local matter. As both the $G_2$ and the SU(2) divisors are smooth, there
are $g=1+\tfrac12\mathcal{S}_9\cdot(\mathcal{S}_9+K_B)$ adjoint matter fields in the $\mathbf{14}$ representation of $G_2$
and $g_{\text{SU}(2)}=1+\tfrac12(-3K_B-\mathcal{S}_7-\mathcal{S}_9)\cdot(-2K_B-\mathcal{S}_7-\mathcal{S}_9)$ adjoints in the $\mathbf{3}$ representation of SU(2), respectively. This yields the 
last two lines of Table \ref{tab:XSU2G2matter}.

For $G_2$, the fundamental representation
$\mathbf{7}$ is in general non-local, as already discussed in
\cite{Grassi:2011hq}.   The multiplicity of this representation is
given\footnote{Thanks to D.\ Morrison for discussions on this point.}
by the difference $g'-g$. Here $g$ is the genus of the $G_2$-divisor $s_9=0$ and $g'$ is the 
genus of the threefold cover\footnote{This is expected as the gauge group 
$G_2$ arises by acting with the outer automorphism $\mathbb{Z}_3$ on
the Dynkin diagram of SO(8).}  of the curve $s_9=0$ with branch points $p$ 
given by the codimension two enhancement points $s_9=\Delta'=0$ with $\Delta'$ given in \eqref{eq:WSFG2SU2} \cite{Grassi:2011hq}.
Using the Riemann-Hurwitz formula for the genus $g'$ of a ramified covering of a 
genus $g$ Riemann surface  \cite{griffiths2014principles},
\beq
	g'=\tfrac12(2+N(2g-2)+\sum_{p}(e_p-1))\,,
\eeq
where $N$ is the degree of the covering,  $p$ are its branch points and $e_p$ denotes the 
ramification index at $p$, we obtain using $N=3$ and $e_p=2$ at all $p$:
\beq\label{eq:g'-g}
	g'-g=(-2 K_B + \mathcal{S}_7 - \mathcal{S}_9)\cdot \mathcal{S}_9\,.
\eeq
This follows as there are $\mathcal{S}_9\cdot [\Delta']=2\mathcal{S}_9\cdot (-3 K_B + \mathcal{S}_7 - 2 \mathcal{S}_9)$ identical branch points $p$ and since $g=1+\tfrac12\mathcal{S}_9\cdot(\mathcal{S}_9+K_B)$. We note that \eqref{eq:g'-g} is precisely the multiplicity in  the second line of Table \ref{tab:XSU2G2matter}. 

The enhancement points
$V_{\text{bf}}=\{s_9=\tilde{s}_1=0\}$ support bifundamental matter. The $(\mathbf{2},\mathbf{7})$
representation is self-conjugate, and thus allows for
half-hypermultiplets; indeed, as encountered in
the context of non-Higgsable clusters \cite{Morrison:2012np}, each
such point supports a half-hypermultiplet in this representation. The number of bifundamentals
is thus given by $\mathcal{S}_9\cdot(-3K_B-\mathcal{S}_7-\mathcal{S}_9)$ yielding the 
first line in Table \ref{tab:XSU2G2matter}.
In addition to supporting bifundamentals, 
at the intersection points $s_9 = \tilde{s}_1 = 0$
there must also
be one additional $(\mathbf{2},\mathbf{1})$ representation at
$s_0=\tilde{s}_1=0$. 
As in the analysis of  \cite{Grassi:2011hq, Morrison:2012np}, this can
be seen by analyzing the matter structure through the monodromy cover;
the $G_2$ can be enhanced to an $SO(7)$, under which the ${\bf 7} +
{\bf 1}$ of $G_2$ combine to a spinor ${\bf 8}$ representation.
Taking into account the SU(2) fundamentals at the points $V_2=\{\tilde{s}_1=4\tilde{s}_2\tilde{s}_4-\tilde{s}_3^2=0\}$ of $I_3$ fibers, we obtain, using \eqref{eq:cubicsections}, 
the third line of Table \ref{tab:XSU2G2matter}.

\subsubsection*{Deformations of $I_2\times I_0^{*\text{ns}}$ Weierstrass models}

Finally, we reverse  our perspective and apply the above results to describe how to deform an 
elliptic fibration with standard $I_2$  and $I_0^{*\text{ns}}$ 
singularities, {\it i.e.}~an F-theory geometry with $\text{SU}(2)\times G_2$ gauge symmetry, to 
a ``Higgsed'' elliptic fibration with only an $I_2$ singularity, {\it i.e.}~an $\text{SU}(2)$ gauge group, but 
codimension two singularities giving rise to the discussed matter in the 
three-index symmetric tensor representation. The idea is to start with the tuned geometry 
specified by the Weierstrass model \eqref{eq:WSFG2SU2} and to 
view the original model defined by $X^{\text{SU}(2)}$ as a deformation thereof. To this end, we 
introduce the deformation parameter
\beq \label{eq:epsilon}
	\epsilon:=s_8-a s_9
\eeq
describing the deviation from the tuning \eqref{eq:G2tuning}. The class of $s_8$, expressed in terms of the classes of the SU(2) and $\text{G}_2$ divisors $\tilde{s}_1$ and $s_9$, respectively, reads
\beq
 [\epsilon]=2[s_9]+[\tilde{s}_1]+2K_B\,,
\eeq
which imposes a minimal degree of $\tilde{s}_1$ and $s_9$ for the deformations $\epsilon$ to exist. In addition, this implies that the  degree of $\epsilon$ is completely fixed if the degrees of  $\tilde{s}_1$ and $s_9$ are given.
Employing the parametrization of the Weierstrass model \eqref{eq:WSFG2SU2}  in terms of the sections $\tilde{s}_1$, $\tilde{s}_2$, $\tilde{s}_3$, $s_4$ and $s_9$ 
as well as the definition of $\epsilon$, we express the deformed Weierstrass model \eqref{eq:WSFSU2} as
\bea \label{eq:WSFdeformed}
f&=&(-\tfrac{1}{3} \tilde{s}_2^2 +\tilde{s}_3\tilde{s}_1) s_9^2+(\tfrac{1}{3} \tilde{s}_2 \tilde{s}_3  -3\tilde{s}_1  \tilde{s}_4) s_9 \epsilon+(\tilde{s}_2 \tilde{s}_4-\tfrac13 \tilde{s}_3^2) \epsilon ^2\,,\\
g&=&(-\tfrac{2}{27} \tilde{s}_2^3  +\tfrac{1}{3} \tilde{s}_2 \tilde{s}_3 \tilde{s}_1- \tilde{s}_4\tilde{s}_1^2)s_9^3+(\tilde{s}_1 (\tilde{s}_2 \tilde{s}_4-\tfrac{2}{3} \tilde{s}_3^2 )+\tfrac{1}{9} \tilde{s}_2^2 \tilde{s}_3)s_9^2 \epsilon\nn\\ &&+(\tfrac{1}{9} \tilde{s}_2 \tilde{s}_3^2 -\tfrac{2}{3} \tilde{s}_2^2 \tilde{s}_4 +\tilde{s}_1\tilde{s}_3 \tilde{s}_4) s_9 \epsilon ^2+(\tfrac{1}{3} \tilde{s}_2 \tilde{s}_3 \tilde{s}_4-\tfrac{2}{27}\tilde{s}_3^3-\tilde{s}_1 \tilde{s}_4^2) \epsilon ^3\,.\nn
\eea
We readily check that  \eqref{eq:WSFdeformed} reduces to \eqref{eq:WSFG2SU2} in the limit $\epsilon\rightarrow 0$. Its $I_2$ singularity is located, in the employed parametrization, at
\beq
	t=- \tilde{s}_1 s_9^3+ \tilde{s}_2 s_9^2 \epsilon - \tilde{s}_3 s_9 \epsilon ^2+ \tilde{s}_4 \epsilon ^3=0
\eeq
with ordinary triple point singularities at $s_9=\epsilon=0$. 

In field theory, the above deformation corresponds to a Higgsing of the  SU(2)$\times G_2$ 
theory. Indeed, we see that the spectrum in Table \ref{tab:XSU2G2matter} exactly reproduces
the SU(2) spectrum in Table \ref{tab:XSU2matter} as
\bea
	&x_\mathbf{4}=x_{(\mathbf{2},\mathbf{7})} + 2 (x_{(\mathbf{1},\mathbf{14})} - 1)\,,\quad x_{\mathbf{3}}= x_{(\mathbf{3},\mathbf{1})} +2x_{(\mathbf{2},\mathbf{7})}+x_{(\mathbf{1},\mathbf{7})}+x_{(\mathbf{1},\mathbf{14})}-1\,,&\nn\\
	&  x_{\mathbf{2}}=2x_{(\mathbf{1},\mathbf{7})} + x_{(\mathbf{2},\mathbf{7})} +x_{(\mathbf{2},\mathbf{1})}\,, &
\eea
where the $-2$ and $-1$ in the multiplicities take into account the fields eaten up by the
massive gauge bosons.
This corresponds to the group theoretical breaking
\beq
	\text{SU}(2)\times G_2\supset \text{SU}(2)^3\longrightarrow \text{SU}(2)\,,
\eeq
where we first embed the regular subgroup SU$(2)^2$ into $G_2$ and then break to SU(2).
The relevant representations branch as
\bea \label{eq:G2SU2->SU2}
 &(\mathbf{1},\mathbf{14})\cong (\mathbf{1},\mathbf{1},\mathbf{3})\oplus  (\mathbf{1},\mathbf{3},\mathbf{1})\oplus  (\mathbf{1},\mathbf{2},\mathbf{4})\longrightarrow \mathbf{3}\oplus 3\cdot\mathbf{1}\oplus 2\cdot \mathbf{4}\,,&\nn
\\
 &(\mathbf{2},\mathbf{7})\cong (\mathbf{2},\mathbf{1},\mathbf{3})\oplus  (\mathbf{2},\mathbf{2},\mathbf{2})\longrightarrow \mathbf{4}\oplus\mathbf{2}\oplus 2\cdot(\mathbf{3}\oplus\mathbf{1})\,,&\nn 
 \\ 
 &(\mathbf{1},\mathbf{7})\cong (\mathbf{1},\mathbf{1},\mathbf{3})\oplus  (\mathbf{1},\mathbf{2},\mathbf{2})\longrightarrow \mathbf{3}\oplus 2\cdot\mathbf{2}\,,&\nn
 \\ 
& (\mathbf{3},\mathbf{1})\cong(\mathbf{3},\mathbf{1},\mathbf{1})\rightarrow \mathbf{3}\,,\qquad (\mathbf{2},\mathbf{1})\cong (\mathbf{2},\mathbf{1},\mathbf{1})\rightarrow \mathbf{2}\,.&
\eea
Here, we denote by $\cong$ the presentation of $\text{SU}(2)\times G_2$ irreducible 
representations as (reducible) representations of its subgroup $\text{SU}(2)^3$.
The embedding of the final SU(2) gauge group into $\text{SU}(2)^3$ is such that 
representations of the middle SU(2) go to multiple copies of singlets
and the tensor product of the representations of the two outer SU(2)'s
is formed, 
{\it i.e.}~$(\mathbf{R},\mathbf{R}',\mathbf{R}'')\rightarrow \text{dim}(\mathbf{R}')\cdot (\mathbf{R}\otimes \mathbf{R}'')$.

The Higgs fields leading to the particular branching \eqref{eq:G2SU2->SU2} transform
in the $\text{SU}(2)\times G_2$-representation $(\mathbf{2},\mathbf{7})$.  There are 17 
vector multiplets before and three after Higgsing. The 14 vector multiplets that get massive in the 
Higgsing transform according to the first line in \eqref{eq:G2SU2->SU2} as 
one $\mathbf{3}$, three singlets $\mathbf{1}$ and two $\mathbf{4}$'s of the final SU(2).
They eat up hypermultiplets in the broken $(\mathbf{2},\mathbf{7})$ in the corresponding representations in the second line of \eqref{eq:G2SU2->SU2}. Thus, for this Higgsing to be possible there have to be four half-hypermultiplets  in the real representation $(\mathbf{2},\mathbf{7})$.\footnote{The number of half-hypers in the $(\mathbf{2},\mathbf{7})$ is given, according to Table \ref{tab:XSU2G2matter}, by $\mathcal{S}_9\cdot(-3K_B-\mathcal{S}_7-\mathcal{S}_9)\geq \mathcal{S}_9\cdot(-2K_B-\mathcal{S}_9)$ for $[s_8]\geq [s_9]$. E.g.~for $B=\mathbb{P}^2$ all models of the form \eqref{eq:WSFG2SU2} have at least 5 half-hypers.} The Higgs VEVs have to be turned on along the singlet components in the second
line of \eqref{eq:G2SU2->SU2}. As just mentioned, three SU(2)-singlet hypermultiplets are eaten 
up by the massive vector multiplets. Thus, also three complex Higgs VEVs have to be fixed by supersymmetry.
It would be interesting to understand this condition explicitly on the
level of D-term constraints in the 6D effective SUGRA theory, which
should describe the full moduli space of the resulting Higgsed theory being parametrized by
all singlets in the breaking \eqref{eq:G2SU2->SU2} with three fields
  fixed by D-flatness.
Note that in the case mentioned above, for example, where on $\P^2$ we
can tune a $G_2$ factor on a line, $[s_9] = 1$, and an SU(2) on a conic,
$[s_1] = 2$, there are
only two half-hypermultiplets in the $(\mathbf{2},\mathbf{7})$
representation, explaining the inability to Higgs the model in this
and other such cases, and correlating with the absence of an appropriate
SU(2) model violating the effectiveness constraints from \eqref{eq:cubicsections}.

\subsection{UnHiggsing SU(2) with $\mathbf{4}$ to $\text{SU}(2)^3$ with trifundamentals}

We conclude with a brief discussion of a different  unHiggsing of  the SU(2) model defined by
F-theory on $X^{\text{SU}(2)}$ leading
to a theory with three $\text{SU}(2)$ gauge algebras on three
different divisors and with a matter spectrum which necessarily has to contain a
trifundamental representation besides the standard
adjoint, fundamental and bifundamental representation. 

The unHiggsing is preformed by imposing that the divisor $t=0$ defined in
\eqref{eq:SU2divisor} factorizes as 
\beq
t=s_4s_8^3-s_3s_8^2s_9+s_2s_8s_9^2-s_1s_9^3\stackrel{!}{=}\prod_{i=1}^3(a_i
s_8+b_i s_9)\,.  
\eeq 
This imposes the obvious constraints of the form
\beq \label{eq:SU2SU2SU2cond} s_4=a_1a_2a_3\,,\!\quad
\!s_3=\!-a_1a_2b_3-a_1a_3b_2-a_2a_3b_1\,,\!\quad\!
s_2=a_1b_2b_3+a_2b_1b_3+a_3b_1b_2\,,\!\quad\! s_1=\!-b_1b_2b_3\,.  \eeq
We note that under this tuning, the Weierstrass model
\eqref{eq:WSFSU2} that is obtained by the special solution
\eqref{eq:simpleSol} develops six singularities of Kodaira type
$I_2$. This is attributed to the fact that the simple solution
overspecializes the complex structure of $X^{\text{SU}(2)}$, leading
to spurious singularities. 

A more general Weierstrass form is obtained
over a UFD using the tuning \eqref{eq:UFDSolution}. In this case, imposing the conditions
\eqref{eq:SU2SU2SU2cond} introduces three singularities of Kodaira
type $I_2$ along the three divisors 
\beq 
t_i:=a_is_8+b_is_9=0\,,
\qquad i=1,2,3\,.  
\eeq 
The resulting Weierstrass model is
algebraically very complex.  Instead of presenting it here, we just mention its key properties. 
A careful analysis of its codimension two singularities reveals that the resulting model
has matter in the fundamental representations w.r.t.~all three SU(2) factors as well as in
all possible bifundamental representations of two SU(2)'s. Most notably, at the  
codimension two locus $s_8=s_9=0$ the three SU(2) divisors $t_i=0$ intersect. 
Employing the fact that the Weierstrass model is not of the standard $I_2$ form following from 
Tate's algorithm, it can be argued that there is trifundamental matter located
at these points. This is also required by the Higgsing back to the original SU(2) model
specified by $X^{\text{SU}(2)}$. 

We will return to analyzing $\text{SU}(2)^3$ models with
trifundamental matter and their (un-)Higgsings in future work
\cite{kmrt}.


\section{Conclusions}
\label{sec:conclusions}

In this paper we have presented an explicit construction of a class of
Weierstrass models that realize matter in the three-index symmetric
({\bf 4}) representation of SU(2).  For 6D F-theory models, this
matter is localized at triple point singularities in the curve $C$
carrying the gauge group.  Such singularities have a contribution $g_a
= 3$ to the arithmetic genus of $C$, matching with the formula
(\ref{eq:genus-contribution}) and the conjectured interpretation of
this formula in \cite{Kumar:2010am}.  To our knowledge, this
represents the first explicit realization 
 in the
  F-theory literature
of any matter representation
with a genus contribution $g > 1$ through a Weierstrass model.

In the Weierstrass models studied here the gauge group lives on a
curve of the form $t =A \xi^3 + B \xi^2 \eta + C \xi \eta^2 + D
\eta^3$, where the triple point singularities are found at the locus
of points satisfying $\xi = \eta = 0$.  This is closely parallel to
the framework of \cite{Cvetic:2015ioa, Anderson:2015cqy}, where
two-index symmetric matter was found to live on curves of the form $t
=A \xi^2 + B \xi \eta + C \eta^2$.  Here, as in those papers, the
vanishing of the discriminant $\Delta$ to order $N$ for an $I_N$
singularity depends on the singular structure of
$t$, and the Weierstrass model does not take the simple form that
follows when one starts from the general Tate model for an $I_N$
singularity on a general divisor $t$ and transforms to Weierstrass
form.  This matches with the analysis of \cite{Anderson:2015cqy}, in
which transitions between theories with different matter content were
studied.  It was found there that for 6D theories, a transition
between two models with distinct matter representations and a given
gauge group occurs when the model passes through a superconformal
fixed point.  Indeed, by continuity it seems impossible to change
matter representations without such a transition when the gauge group
is kept fixed.  Thus, for example, tuning a Tate type model with an
SU($N$) gauge group on a smooth curve $C$ and then taking a singular
limit of $C$ cannot change the matter content, so the full genus
contribution must still come from adjoint matter in any model where
the Weierstrass model comes from the generic Tate $I_N$ form.  This
explains the necessity for the remarkable algebraic structure involved
in the realizations of the symmetric matter representations found in
this and previous works.

Another remarkable feature of the analysis here is that the
Weierstrass form of the U(1) models of \cite{Klevers:2014bqa} that we have used does not seem to
fit in the general classification given in \cite{Morrison:2012ei}.  In
that paper a general argument was given suggesting that any F-theory
model with an Abelian factor should have a Weierstrass description of
the form
\begin{equation}
 y^2 = x^3+ (c_1c_3-b^2c_0- \frac{1}{3}c_2^2 ) x
+ (c_0c_3^2 -\frac{1}{3}c_1c_2c_3 + \frac{2}{27}c_2^3 - \frac{2}{3}b^2c_0c_2
+\frac{1}{4}b^2c_1^2) \,.
\label{eq:mp-u1}
\end{equation}
The Weierstrass models for U(1) theories with charge $q = 3$ matter we
consider here, do not, however, seem to take this form \cite{Klevers:2014bqa}.  In fact, we
would have a problem if they did.  It was argued in
\cite{Morrison:2012ei, Morrison:2014era} that in any U(1) model of the
form (\ref{eq:mp-u1}), taking $b \rightarrow 0$ gives an unHiggsing to
an SU(2) model.  The resulting SU(2) model, however is always in the
form that follows by starting with a generic Tate $I_2$ construction, with the
SU(2) realized on the divisor $\{c_3 = 0\}$, and transforming to
Weierstrass form.  It seems then 
from the discussion above and the analysis of
\cite{Anderson:2015cqy}
that any
such SU(2) can only have $g_R > 0$ matter coming from adjoint
  representations and cannot include exotic matter such as 
  three-index symmetric matter representations.  Thus, the existence
  of these constructions seems to suggest that there must be a more
  general class of U(1) models than those constructed in
\cite{Morrison:2012ei}.  We can understand this further by considering
that in \cite{Morrison:2012ei} the form (\ref{eq:mp-u1}) arose from a
situation where the extra section had an explicit description through
\begin{equation}
[x, y, z] =[c_3^2 -\frac{2}{3} b^2 c_2, -c_3^3 + b^2
  c_2c_3-\frac{1}{2}b^4c_1, b] \,.
\label{eq:QinMP}
\end{equation}
Comparing to the expressions for the section $[x_1, y_1, z_1]$ in
Appendix~\ref{app:RepInWSF}, we find that in our case there is a
similar description, where identifying $b \equiv z_1 =s_7 s_8^2 - s_6 s_8
s_9 + s_5 s_9^2$ the section can be described in the form
\begin{equation}
[x, y, z] =[c_3^2 -\frac{2}{3} b c_2, -c_3^3 + b
  c_2c_3-\frac{1}{2} b^2 c_1, b] \,.
\label{eq:QfordP1}
\end{equation}
Understanding better how to construct more general classes of U(1)
models with higher charges that allow unHiggsing to non-Abelian SU(2)
models with exotic matter representations may shed light on the
general construction of Weierstrass models where gauge groups are
realized on singular divisors. A natural starting point, for example, is the complete 
intersection $\text{U}(1)^3$ model in \cite{Cvetic:2013qsa}

This paper has presented a novel and specific example of a rather remarkable
geometric and algebraic structure that can arise in F-theory, adding
to the small set of explicit classes of Weierstrass models known that
realize exotic matter representations.  There are many ways in which
it would be interesting to expand on these developments, both in terms
of this and other specific realizations and in terms of more general
theoretical structures.

For the specific class of representations studied here, namely the
{\bf 4} of SU(2), it would be interesting to analyze the dual
heterotic models in cases with a smooth heterotic dual, as was done
for the two-index symmetric representation of SU(3) in
\cite{Anderson:2015cqy}.  Also following the lines of
\cite{Anderson:2015cqy}, it seems that analogous constructions to those
found here can be realized explicitly through exotic matter
transitions in a further unHiggsed non-Abelian theory; results on this
will be presented elsewhere \cite{kmrt}.\footnote{We thank Nikhil Raghuram for
discussions related to this issue.} 

In principle, the methods used here could be used to construct larger
exotic SU($N$) representations.
To follow the same logic as that presented here for higher-dimensional
representations of SU(2), for example, we would need to identify
models with U(1) gauge fields and matter fields transforming under
representations of charge $q > 3$.  
More generally, it would be desirable to address the general challenge
of classifying the algebraic structures that can be used in the
Weierstrass model to construct general gauge groups over singular
divisors, and to bring together algebraic, geometric, and field theory
understandings of these more exotic matter representations along with
their Higgsings and unHiggsings to theories with Abelian or
higher-rank non-Abelian gauge theories.  This seems like a rich arena
for exploration, with highly intricate and nontrivial structure in the
Weierstrass models encoding these features, and we anticipate that
further study of these questions will lead to additional novel
results and increased understanding.  Finally, getting
a systematic handle on the types of codimension two singularities that
can be realized in Weierstrass models for elliptically fibered Calabi-Yau manifolds would be
an important step towards completing the systematic classification of
 such geometries \cite{Morrison:2012np,
Morrison:2012js, Taylor:2012dr, Martini:2014iza, Johnson:2014xpa,
Taylor:2015isa}.

\subsubsection*{Acknowledgments}

We would like to thank Lara Anderson, Mirjam Cveti\v{c}, Clay Cordova, Antonella Grassi, James Gray, Albrecht Klemm, Noppadol Mekareeya, David Morrison, and
Nikhil Raghuram for
helpful discussions.
We are grateful to  the 2015 Summer Program on “F-Theory at the interface of particle physics and mathematics” at the Aspen Center for Physics for hospitality during the course of the project. 
The work of WT
 was supported by the DOE under contract 
\#DE-SC00012567.

\appendix

\section{Representation in Tate and Weierstrass form}
\label{app:RepInWSF}

Here we present the explicit expressions for the Weierstrass model of the $dP_1$-elliptic fibration
$X$ in \eqref{eq:PF3}. We refer the reader to \cite{Klevers:2014bqa} for more details.

We apply Nagell's algorithm to the cubic \eqref{eq:PF3} with
respect to the point $\hat{c}_0\cap \mathcal{E}$ to obtain a birational map to its WSF.
We determine the functions $f$, $g$ of this WSF to be given by
\footnotesize
\bea
\label{eq:fgcubic}
f\!\!&\!\!=\!\!&\!\!\tfrac{1}{48} \left(24 \left(2 \left(s_2 s_4 s_8^2+s_1 \left(s_7^2-3 s_4 s_9\right) s_8+s_5 \left(s_4 s_5+s_2 s_7\right) s_9+s_3 \left(s_5 s_7 s_8+s_9 \left(s_2 s_8+s_1 s_9\right)\right)\right)\right.\right.\\
&-\!\!&\!\!\left.\left.s_6 \left(s_4 s_5 s_8+s_2 s_7 s_8+\left(s_3 s_5+s_1 s_7\right) s_9\right)\right)-\left(s_6^2-4 \left(s_5 s_7+s_3 s_8+s_2 s_9\right)\right){}^2\right)\nn\\
g\!\!&\!\!=\!\!&\!\!\tfrac{1}{864} \left(\left(s_6^2-4 \left(s_5 s_7+s_3 s_8+s_2 s_9\right)\right){}^3-36 \left(2 \left(s_2 s_4 s_8^2+s_1 \left(s_7^2-3 s_4 s_9\right) s_8+s_5 \left(s_4 s_5+s_2 s_7\right) s_9\right.\right.\right.\nn\\ 
&+\!\!&\!\!\left.\left.\left.s_3 \left(s_5 s_7 s_8+s_9 \left(s_2 s_8+s_1 s_9\right)\right)\right)-s_6 \left(s_4 s_5 s_8+s_2 s_7 s_8+\left(s_3 s_5+s_1 s_7\right) s_9\right)\right) \left(s_6^2-4 s_5 s_7-4s_3 s_8\right.\right.\nn\\ 
&-\!\!&\!\!\left.\left.4s_2 s_9\right)
+216 \left(\left(s_2^2-4 s_1 s_3\right) s_7^2 s_8^2+s_4^2 \left(s_5^2-4 s_1 s_8\right) s_8^2-2 s_7 \left(s_2 \left(s_3 s_5+s_1 s_7\right)-2 s_1 s_3 s_6\right) s_9 s_8\right.\right.\nn\\
&+\!\!&\!\!\left.\left.\left(\left(s_3 s_5-s_1 s_7\right){}^2-4 s_1 s_3^2 s_8\right) s_9^2+2 s_4 \left(-2 s_1^2 s_9^3+2 \left(s_1 s_5 s_6-s_2 \left(s_5^2-2 s_1 s_8\right)\right) s_9^2\right.\right.\right.\nn
\\
~&-\!\!&\!\!\left.\left.\left.s_8 \left(2 s_8 s_2^2-2 s_5 s_6 s_2+2 s_1 s_6^2+s_1 s_5 s_7+s_3 \left(s_5^2-4 s_1 s_8\right)\right) s_9+\left(2 s_1 s_6-s_2 s_5\right) s_7 s_8^2\right)\right)\right)\nn
\eea
\normalsize
We observe that
there is no factorization of the discriminant $\Delta$ following from $f$ and $g$ indicating the absence of
codimension one singularities and a non-Abelian gauge group.

Furthermore, we plug the  coordinates of the rational section \eqref{eq:s0F3} into
this map to obtain its coordinates in  WSF,
\footnotesize
\bea\label{eq:WSFQ1F3}
	z_{1}\!\!&\!\!=\!\!&\!\!s_7 s_8^2 - s_6 s_8 s_9 + s_5 s_9^2\,,\!\!\!\!\!\!\!\!\\
	 x_{1}\!\!&\!\!=\!\!&\!\! \tfrac{1}{12} \left(12 s_1^2 s_9^6+4 \left(2 s_2 \left(s_5^2-3 s_1 s_8\right)-3 s_1 s_5 s_6\right) s_9^5+\left(\left(s_6^2-4 s_5 s_7\right) s_5^2+12 \left(s_2^2+2 s_1 s_3\right) s_8^2-4 \left(4 s_3 s_5^2\right.\right.\right.\nn\\ 
	 \!\!&\!\!+\!\!&\!\!\!\!\!\!\left.\left.\left.s_2 s_6 s_5-3 s_1 \left(s_6^2+2 s_5 s_7\right)\right) s_8\right) s_9^4-2 s_8 \left(-4 \left(s_6 s_7+3 s_4 s_8\right) s_5^2+\left(s_6^3-10 s_3 s_8 s_6+4 s_2 s_7 s_8\right) s_5\right.\right.\nn \\
	  \!\!&\!\!+\!\!&\!\!\!\!\!\!\left.\left.2 s_8 \left(9 s_1 s_6 s_7+6 s_1 s_4 s_8+s_2 \left(s_6^2+6 s_3 s_8\right)\right)\right) s_9^3+s_8^2 \left(s_6^4-2 s_5 s_7 s_6^2-8 s_5^2 s_7^2+12 \left(s_3^2+2 s_2 s_4\right) s_8^2\right.\right.\nn\\ 
	  \!\!&\!\!-\!\!&\!\!\!\!\!\!\left.\left.4 \left(9 s_4 s_5 s_6-s_7 \left(5 s_2 s_6+6 s_1 s_7\right)+s_3 \left(s_6^2+2 s_5 s_7\right)\right) s_8\right) s_9^2-2 s_8^3 \left(12 s_3 s_4 s_8^2+2 \left(s_7 \left(s_3 s_6+4 s_2 s_7\right)\right.\right.\right.\nn\\
	   \!\!&\!\!-\!\!&\!\!\!\!\!\!\left.\left.\left.3 s_4 \left(s_6^2+2 s_5 s_7\right)\right) s_8+s_6 s_7 \left(s_6^2\!-\!4 s_5 s_7\right)\right) s_9+s_8^4 \left(\left(s_6^2\!-\!4 s_5 s_7\right) s_7^2+4 \left(2 s_3 s_7\!-\!3 s_4 s_6\right) s_8 s_7+12 s_4^2 s_8^2\right)\right)\,,\!\!\!\!\!\!\!\!\nn\\
	  y_{1}\!\!&\!\!=\!\!&\!\!\frac{1}{2} \left(2 s_1^3 s_9^9+s_1 \left(2 s_2 \left(s_5^2-3 s_1 s_8\right)-3 s_1 s_5 s_6\right) s_9^8+\left(\left(s_3 s_5^2-s_2 s_6 s_5+s_1 \left(s_6^2-s_5 s_7\right)\right) s_5^2\right.\right.\nn\\ 
	  &+\!\!&\!\!\left.\left.
	  6 s_1 \left(s_2^2+s_1 s_3\right) s_8^2+\left(-2 s_2^2 s_5^2+2 s_1 s_2 s_6 s_5+s_1 \left(3 s_1 \left(s_6^2+2 s_5 s_7\right)-4 s_3 s_5^2\right)\right) s_8\right) s_9^7\right.\nn\\ 
	  &-\!\!&\!\!\left.\left. s_8 \left(2 \left(s_2^3+6 s_1 s_3 s_2+3 s_1^2 s_4\right) s_8^2-(s_5 s_6 s_2^2+\left(6 s_3 s_5^2-4 s_1 \left(s_6^2+2 s_5 s_7\right)\right) s_2+s_1 \left(6 s_4 s_5^2+2 s_3 s_6 s_5\right.\right.\right.\right.\nn\\ 
	  &-\!\!&\!\!\left.\left.\left.\left. 9 s_1 s_6 s_7\right)\right) s_8
	  +s_5 \left(3 s_4 s_5^3+2 s_3 s_6 s_5^2-3 s_2 s_7 s_5^2-2 s_2 s_6^2 s_5+s_1 s_6 s_7 s_5+2 s_1 s_6^3\right)\right) s_9^6+s_8^2 \left(s_1 s_6^4\right.\right.\nn\\ 
	  &-\!\!&\!\!\left.\left.s_2 s_5 s_6^3+s_3 s_5^2 s_6^2+
	  7 s_1 s_5 s_7 s_6^2+9 s_4 s_5^3 s_6-8 s_2 s_5^2 s_7 s_6+s_1 s_5^2 s_7^2+6 \left(s_3 \left(s_2^2+s_1 s_3\right)+2 s_1 s_2 s_4\right) s_8^2
\right.\right.\nn\\ 
	  &-\!\!&\!\!\left.\left.	  s_3 s_5^3 s_7+
	  \left(s_2^2 s_6^2\!-\!4 s_3^2 s_5^2\!-\!8 s_2 s_4 s_5^2\!-\!6 s_1 s_4 s_6 s_5+6 s_1^2 s_7^2+2 s_2 \left(s_2 s_5+7 s_1 s_6\right) s_7+s_3 \left(2 s_1 \left(s_6^2+2 s_5 s_7\right)
	  \right.\right.\right.\right.\nn\\ 
	  &-\!\!&\!\!\left.\left.\left.\left. 6 s_2 s_5 s_6\right)\right) s_8\right) s_9^5
	  -s_8^3 \left(s_8 \left(6 s_2 s_8-5 s_5 s_6\right) s_3^2-5 s_6 s_7 \left(s_5^2-2 s_1 s_8\right) s_3+5 s_7 \left(s_6 s_8 s_2^2+2 s_1 s_7 s_8 s_2
\right.\right.\right.\nn\\ 
	  &-\!\!&\!\!\left.\left.	 \left. s_2 s_5 \left(s_6^2+s_5 s_7\right)
	 + s_1 s_6 \left(s_6^2+2 s_5 s_7\right)\right)+s_4 \left(5 \left(2 s_6^2+s_5 s_7\right) s_5^2-10 \left(s_3 s_5+s_2 s_6\right) s_8 s_5\right.\right.
\right.\nn\\ 
	  &+\!\!&\!\!\left.\left.\left.6 \left(s_2^2+2 s_1 s_3\right) s_8^2\right)\right) s_9^4+
	  s_8^4 \left(2 \left(s_3^3+6 s_2 s_4 s_3+3 s_1 s_4^2\right) s_8^2-\left(6 s_4^2 s_5^2+s_3^2 s_6^2-4 \left(s_2^2+2 s_1 s_3\right) s_7^2
\right.\right.\right.\nn\\ 
	  &+\!\!&\!\!\left.\left.	  \left.2 s_3 \left(s_3 s_5-3 s_2 s_6\right) s_7+
	  2 s_4 \left(s_2 s_6^2+7 s_3 s_5 s_6-3 s_1 s_7 s_6+2 s_2 s_5 s_7\right)\right) s_8+5 \left(s_4 s_5 s_6 \left(s_6^2+2 s_5 s_7\right)
\right.\right.\right.\nn\\ 
	  &+\!\!&\!\!\left.\left.	  \left.\left.s_7 \left(s_7 \left(2 s_1 s_6^2-s_2 s_5 s_6+s_1 s_5 s_7\right)-
	  s_3 s_5 \left(s_6^2+s_5 s_7\right)\right)\right)\right) s_9^3-s_8^5 \left(3 s_8 \left(2 s_2 s_8-3 s_5 s_6\right) s_4^2+\left(s_6^4
\right.\right.\right.\right.\nn\\ 
	  &+\!\!&\!\!\left.\left.	\left.  \left(7 s_5 s_7-4 s_3 s_8\right) s_6^2+2 s_2 s_7 s_8 s_6+s_5^2 s_7^2-
	  8 s_3 s_5 s_7 s_8+6 s_8 \left(s_8 s_3^2+s_1 s_7^2\right)\right) s_4+s_7 \left(s_6 s_8 s_3^2-\left(s_6^3\right.\right.\right.
\right.\nn\\ 
	  &+\!\!&\!\!\left.\left.\left.\left. 8 s_5 s_7 s_6-6 s_2 s_7 s_8\right) s_3+s_7 \left(9 s_1 s_6 s_7+s_2 \left(s_6^2-s_5 s_7\right)\right)\right)\right) s_9^2+
	  s_8^6 \left(3 s_8 \left(-s_6^2-2 s_5 s_7+2 s_3 s_8\right) s_4^2
\right.\right.\nn\\ 
	  &+\!\!&\!\!\left.\left.	  \left.s_7 \left(2 s_6^3+s_5 s_7 s_6-2 s_3 s_8 s_6+4 s_2 s_7 s_8\right) s_4+s_7^2 \left(2 s_8 s_3^2-2 s_6^2 s_3-3 s_5 s_7 s_3+
	  3 s_1 s_7^2+2 s_2 s_6 s_7\right)\right) s_9\right.\right.\nn
	  \eea
\bea
	  &+\!\!&\!\!\left.\left.	  \left. s_8^7 \left(-2 s_8^2 s_4^3+3 s_6 s_7 s_8 s_4^2+s_7^2 \left(-s_6^2+s_5 s_7-2 s_3 s_8\right) s_4+s_7^3 \left(s_3 s_6-s_2 s_7\right)\right)\right)\right.\right.\,.\phantom{................................}\nn
\eea
\normalsize

The Weierstrass form  \eqref{eq:fgcubic} can be obtained from
a Tate model with the following Tate coefficients \cite{Klevers:2014bqa}:
\footnotesize
\bea \label{eq:Tate}
a_1\!\!&\!\!=\!\!&\!\!\!\!s_6\,,\quad a_2=-s_5 s_7-s_3 s_8-s_2 s_9,\quad a_3=-\left(s_4 s_5+s_2 s_7\right) s_8-\left(s_3 s_5+s_1 s_7\right) s_9\,,\!\!\!\!\\
 a_4\!\!&\!\!=\!\!&\!\!\!\! s_1 s_3 s_9^2+\left(s_2 \left(s_5 s_7+s_3 s_8\right)+s_4 \left(s_5^2-3 s_1 s_8\right)\right) s_9+s_8 \left(s_1 s_7^2+s_3 s_5 s_7+s_2 s_4 s_8\right)\,,\nn\\
  a_6\!\!&\!\!=\!\!&\!\!\!\!-s_1^2 s_4 s_9^3-\left(s_2 s_4 \left(s_5^2-2 s_1 s_8\right)+s_1 \left(s_3 \left(s_5 s_7+s_3 s_8\right)-s_4 s_5 s_6\right)\right) s_9^2-s_8 \left(s_4 s_8 s_2^2+\left(s_1 s_7^2-s_4 s_5 s_6\right) s_2\right.\nn\\
 \!\!&\!\!+\!\!&\!\!\!\!\left.s_1 s_4 \left(s_6^2\!+\!s_5 s_7\right)\!+\!s_3 \left(\left(s_2 s_5\!-\!s_1 s_6\right) s_7+s_4 \left(s_5^2\!-\!2 s_1 s_8\right)\right)\right) s_9\!-\!s_8^2 \left(s_2 s_4 s_5 s_7\!+\!s_1 \left(s_8 s_4^2\!+\!s_3 s_7^2\!-\!s_4 s_6 s_7\right)\right)\,.\nn \!\!\!\!\!\!\!\!\!\!\!\!
\eea
\normalsize


\bibliographystyle{utphys}	
\bibliography{ref}

\end{document}